\documentclass[10pt,onecolumn,journal]{IEEEtran}
\usepackage[pass,letterpaper]{geometry} 

\usepackage{cite}

\ifCLASSINFOpdf
\else
\fi
\usepackage{indentfirst}
\usepackage{amsfonts}
\usepackage{amsmath,amsthm,amssymb}
\usepackage{graphicx}
\usepackage{color}
\usepackage{geometry}
\usepackage[style]{fncychap}
\usepackage{extarrows}
\usepackage{tikz}
\usepackage{multirow}
\usepackage{mathrsfs}
\usepackage{pifont}
\usepackage{booktabs}
\usepackage{verbatim}
\usepackage{empheq}
\usepackage{colortbl}
\usepackage{setspace}
\usepackage{epstopdf}
\usetikzlibrary{patterns}

\usetikzlibrary{arrows,shapes}
\usetikzlibrary{shadows}

\newtheorem{Definition}{Definition}

\newtheorem{Lemma}{Lemma}
\newtheorem{Theorem}{Theorem}
\newtheorem{Remark}{Remark}
\newtheorem{Corollary}{Corollary}

\normalsize

\begin{document}

\title{Effects of Feedback on the One-sided Secrecy of Two-way Wiretap through Multiple Transmissions}

\author{\IEEEauthorblockN{Chao Qi\IEEEauthorrefmark{1}\IEEEauthorrefmark{2},
		Yanling Chen\IEEEauthorrefmark{2},
		A. J. Han.Vinck\IEEEauthorrefmark{2}\IEEEauthorrefmark{3} 
		and Xiaohu Tang\IEEEauthorrefmark{1} 
	}
	
	\IEEEauthorblockA{\IEEEauthorrefmark{1}Information Security and National Computing Grid Laboratory,
		Southwest Jiaotong University, Chengdu, China\\ E-mail: chaoqi@my.swjtu.edu.cn, xhutang@home.swjtu.edu.cn}
	\IEEEauthorblockA{\IEEEauthorrefmark{2}Institute of Digital Signal Processing, University of Duisburg-Essen, Germany.
		\IEEEauthorrefmark{3}University of Johannesburg, South Africa}
	E-mail: yanling.chen@uni-due.de, han.vinck@uni-due.de
}

\maketitle

\vspace{5mm}

\begin{abstract}
	In this paper, the one-sided secrecy of two-way wiretap channel with feedback is investigated, where the confidential messages of one user through multiple transmissions is guaranteed secure against an external eavesdropper. 
	For one thing, one-sided secrecy satisfies the secure demand of many practical scenarios. For another, the secrecy is measured over many blocks since the correlation between eavesdropper's observation and the confidential messages in successive blocks, instead of secrecy measurement of one block in previous works. 	
	Thus, firstly, an achievable secrecy rate region is derived for the general two-way wiretap channel with feedback through multiple transmissions under one-sided secrecy. 
	Secondly, outer bounds on the secrecy capacity region are also obtained. 
	The gap between inner and outer bounds on the secrecy capacity region is explored via the binary input two-way wiretap channels. 
	Most notably, the secrecy capacity regions are established for the XOR channel. 
	Furthermore, the result shows that the achievable rate region with feedback is larger than that without feedback. Therefore,  the benefit role of feedback is precisely characterized for two-way wiretap channel with feedback under one-sided secrecy.	
\end{abstract}
\begin{IEEEkeywords}
two-way wiretap channel, secrecy capacity, feedback, achievable secrecy rate region, one-sided secrecy 
\end{IEEEkeywords}

\IEEEpeerreviewmaketitle

	\section{Introduction}	
	 \subsection{The feedback of wiretap channel}
 	The secure communication via a wiretap channel was first studied by Wyner in \cite{wyner1975wire}, where he proved the possibility to achieve a positive secrecy transmission rate without any private key.  
	Wyner studied a noisy degraded broadcast channel and determined its {\it secrecy capacity}, defined to be the maximum transmission rate under a weak secrecy constraint (i.e., the rate of information leaked to the eavesdropper is vanishing).	
	Later, Csisz\'{a}r and K\"{o}rner extended Wyner's work to the general broadcast wiretap channel with a confidential message \cite{csiszar1978broadcast}. 
		
	Notably, in wiretap channel, the positive secrecy rate demands that the legitimate receiver should have a better observation than the eavesdropper does \cite{wyner1975wire,csiszar1978broadcast}.  
	Whereas, in case of a strong eavesdropper who has a better channel than the legitimate receiver does, 
	the works of Maurer, Ahlswede and Csisz\'{a}r have shown that a positive secret key generation rate can be achieved by establishing a public feedback channel \cite{maurer1993secret,ahlswede1993common}. 	
	Thereafter, inspired by the benefit of feedback, a large number of works investigated the role of feedback in various wiretap channel \cite{Ahlswede2006,Chen2005The,Ardestanizadeh2009Wiretap,Permuter2009Capacity,Lai2008The,Gunduz2008Secret}.
	Considering the usage of feedback signal, one class of works sends the feedback back to the channel in order to help confuse the eavesdropper. For example, Ahlswede and Cai \cite{Ahlswede2006} demonstrated that the noiseless feedback from the legitimate receiver to the transmitter can increase the secrecy capacity of the wiretap channel. Lai, El Gamal, and Poor \cite{Lai2008The}, studied the modulo-additive wiretap channel, where Eve receives the modulo-sum of the source signal, the feedback signal, and the noise. They showed that if Bob jams Eve completely, then Alice can send messages securely at the capacity of the channel to Bob.  	
	Another class of works combines the feedback signal and the secret-key exchange mechanism in encoding scheme. For instance, in \cite{Gunduz2008Secret}, feedback signal is utilized as a secret key in one-time pad to derive a higher secrecy rate. 
	  	
	\subsection{Two-way wiretap channel}		
	As a classic multi-users communication model, two-way channel was first studied by Shannon \cite{shannon1961two}, where two users intend to exchange messages with each other.  	
	In the presence of an external eavesdropper, this communication scenario is modeled as the two-way wiretap channel. 
	Tekin and Yener firstly studied the Gaussian two-way wiretap channel \cite{tekin2008general, tekin2010correction}, where an achievable secrecy rate region was derived and the cooperative jamming was proposed to increase the achievable secrecy rate. 	
	The affect of feedback in two-way wiretap channel was investigated in Gaussian two-way wiretap channel \cite{he2013role} and in the general two-way wiretap channel \cite{el2013achievable,pierrot2011strongly}. 
	For Gaussian two-way wiretap channel, the previous received messages was used as feedback and sent into the forward channel to confuse the eavesdropper \cite{he2013role}. Through this method, the achievable secrecy rate is improved than that in   \cite{tekin2008general}. 
	For the general two-way wiretap channel, achievable secrecy rate regions were derived in \cite{el2013achievable} and \cite{pierrot2011strongly} respectively, both of which took the previous messages as feedback and used feedback signal in a secrect-key exchange mechanism to keep the messages confidential. Specially, the work \cite{pierrot2011strongly} studied the two-way wiretap channel under a {\em strong} secrecy constraint, where the amount of information leaked to the eavesdropper, rather than the leakage rate is required to vanish in the limit of the number of channel uses. 
	 
	So far, these works on the two-way wiretap channel focus on the role of feedback with {\em joint secrecy}, where both of the two confidential messages are kept secure. 	
 	However, in many practical communication scenarios, the secrecy requirement may not as strong as the joint secrecy.
 	For instance, in a question-answer communication scenario, two legitimate users communicate with each other over a two-way  channel in presence of an external eavesdropper. Usually, one terminal sends a public question and the other gives a private answer.
 	When considering a weaker secrecy constraint than the joint secrecy, we can only keep the answer confidential from the eavesdropper, instead both the public question and private answer.  
 	This scenario can be modeled as a two-way wiretap channel with one-sided secrecy.
 	Although one-sided secrecy constraint is by definition weaker than the joint secrecy, it nevertheless provides an acceptable secrecy for many practical communication scenarios.   
 	 	
	From previous works, it is known that both cooperative jamming and feedback have been highlighted to increase the secure communication rates in the two-way wiretap channel. 
	In addition, with the full-duplex destination, regarding that the two users' observations are correlated to two users' messages, each legitimate user can generate the feedback signal from part of the received messages with the knowledge of its own message, while the eavesdropped can not gain more information because of the security of feedback.
	Hence, it is interesting to study the impact of cooperative jamming and feedback in two-way wiretap channel under the practical one-sided secrecy constraint. 
	
	Moreover, in this paper by block Markov coding the previous messages of one user is regarded as feedback for the other user, further encrypting the confidential message in one-time pad scheme. Since the channel output is related to both users' messages, such that for one transmission the channel output is not only related to the massages of this transmission, but also related to the messages of previous transmission. 
	Therefore, the information leakage to the eavesdropper should be measured over several transmissions. 
	However, in previous studies of two-way wiretap channel with feedback, the information leakage is investigated for one transmission to explore the role of feedback.	
	Despite the discussion above, the role of feedback in two-way wiretap involving  several transmissions is a problem to be solved.

	\subsection{Contributions of This Paper and Organization}		
	
	Motivated by the meaning of one-sided secrecy and feedback, we investigate the one-sided secrecy over $b$ transmissions of the two-way wiretap channel, which requires the information leakage of the confidential message over $b$ transmissions to vanish. 	
	Under this secrecy constraint, an achievable secrecy rate region for the two-way wiretap is derived by the block Markov coding and the cooperative jamming mechanism.
	Further, the achievable secrecy rate region is explored in binary-input channels. Notably, the secrecy capacity region is fully characterized for the XOR channel. The result shows that the achievable secrecy rate region is larger than that without feedback. Hence, the benefit role of feedback is precisely characterized through multiple transmissions for two-way wiretap channel with one-sided secrecy.
	
	The rest of the paper is organized as follows. In Section \ref{Sec_Model}, we provide the general system model together with the
	preliminary definitions which will be utilized throughout the paper.
	An achievable one-sided secrecy rate region and an outer bound on the one-sided secrecy capacity region are proposed in Section \ref{Sec_Cooperative}.	 
	These results are explored in the binary input two-way wiretap channel in Section \ref{Sec: BiCase}. 
	To enhance the flow of the paper, the detailed proofs are collected in the appendices. 
	
	\section{System Model}\label{Sec_Model}
	\subsection{System Model}
	%
	%
	%
	%

	\begin{figure}[!htbp]
		\centering
		\includegraphics[width=0.65\textwidth]{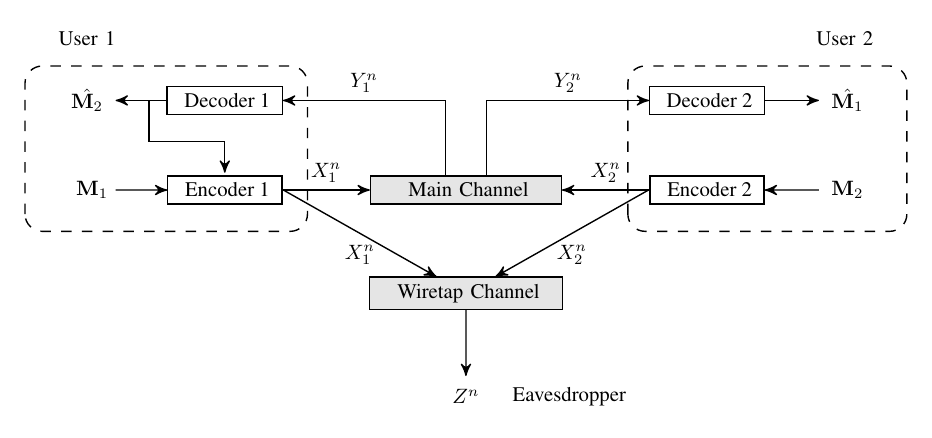}
		\caption{Two-way wiretap channel with an external eavesdropper.}
		\label{Fig: TWWT_channel_Joint}
	\end{figure}   
	
	The full-duplex two-way wiretap channel is shown in Fig. \ref{Fig: TWWT_channel_Joint}, where two legitimate users intend to exchange messages with each other in the presence of an external eavesdropper, and each of users is equipped with a transmitter and a receiver that can operate simultaneously. 
	In this paper, we focus on the one-sided secrecy, where only the messages from the user is kept confidential from the eavesdropper.
	Without loss of generality, assume that only the message from the legitimate user 1 should be kept secure, while the message from user 2 is an open message. 	
	In this paper, the propose Markov block coding scheme involves $b$ transmissions, during which the confidential messages are sent from $2$nd to the $b$-th transmission. We call each transmission one block.
	Before introduce the system, we first introduce the notations and terminologies used later.
	\begin{itemize}
		\item $\mathcal{C}_1$, $\mathcal{C}_2$ are the codebooks of user 1 and user 2, respectively. 
		\item $\mathbf{{M}_1}(j)=( M_{1,1}(j), M_{1,2}(j), \cdots, M_{1,n}(j))$ denotes the $n$ length message of user 1 in the $j$-th block. 	
		Correspondingly, $(\mathbf{M}_1)_j^b$ represents the $j, j+1, \cdots, b$-th message from User 1, and $(\mathbf{M}_1)^j$ represents the $j$ messages of user 1.
		\item Similarly, $\mathbf{M}_2(j)=( M_{2,1}(j), M_{2,2}(j), \cdots, M_{2,n}(j))$ denotes the $n$ length message of user 2 in the $j$-th block. $(\mathbf{M}_2)_j^b$ represents the $j, j+1, \cdots, b$-th message of User 2, and $(\mathbf{M}_2)^j$ represents the $j$ messages of user 2.		
		\item $\mathbf{Z}^j=( \mathbf{Z}(1), \mathbf{Z}(2), \cdots, \mathbf{Z}(j) )$ denotes the messages received by the eavesdropper from $j$ blocks, where \\ 
		$\mathbf{Z}(j)=( Z_1(j), Z_2(j), \cdots, Z_n(j) )$ is the $n$ length message received in the $j$-th block.
	\end{itemize}

	Suppose $\mathcal{M}_{1}$, $\mathcal{M}_{2}$ to be the message sets of user 1 and user 2, respectively; $\mathcal{X}_1$, $\mathcal{X}_2$ to be the finite channel input alphabets at user 1 and user 2, respectively; $\mathcal{Y}_1$, $\mathcal{Y}_2$, $\mathcal{Z}$ are the channel output alphabets at user 1, user 2 and the eavesdropper, respectively. 
	The discrete memoryless two-way wiretap channel is characterized by the transition probability distribution $p(y_1,y_2,z|x_1,x_2)$, where $x_1\in \mathcal{X}_1$, $x_2\in \mathcal{X}_2$ are the channel inputs from user 1 and 2; $y_1\in \mathcal{Y}_1$, $y_2\in \mathcal{Y}_2$ and $z\in \mathcal{Z}$ are channel outputs at user 1, user 2 and the eavesdropper. 
	
	Suppose in block $j-1$, the channel output at user 1 and user 2 are $Y^n_{1}(j-1)$ and $Y^n_{2}(j-1)$, respectively.
	User 1 decodes $Y^n_{1}(j-1)$ and obtains the messages from user 2, part of which was taken as the feedback signal for next block denoted by $Y^n_f(j-1)$.
	In block $j$, the legitimate user 1 wants to transmit a confidential message $\mathbf{M}_{1}\in \mathcal{M}_{1}$ to user 2. According to the feedback $Y^n_f(j-1)$ and $\mathbf{M}_{1}$, the corresponding codeword $X^n_1\in \mathcal{X}_1$ is chosen and sent at a secure rate $\overline{R}_{1s}=\frac{1}{n}H(\mathbf{M}_1)$. 
	The legitimate user 2 wants to transmit a non-secure message $\mathbf{M}_2(j)\in \mathcal{M}_{2}$ to user 1. The corresponding codeword $X^n_2\in \mathcal{X}_2$ is sent at a transmission rate $R_{2}=\frac{1}{n}H(\mathbf{M}_2)$. 
	
	
	A $(2^{n\overline{R}_{1s}},2^{nR_2},n)$ code for the two-way channel consists of the following.
	\begin{itemize}
		\item Two independent message sets $\mathcal{M}_{1}=\{1,2,\ldots,2^{n\overline{R}_{1s}}\}$ and $\mathcal{M}_2=\{1,2,\ldots,2^{nR_2}\}$.
		\item For each transmission, two messages: $\mathbf{M}_{1}$ and $\mathbf{M}_2$ are independent and uniformly distributed over $\mathcal{M}_{1}$ and $\mathcal{M}_2$, respectively. 
		\item Two stochastic encoders, $f_1$ at user 1: $(\mathbf{M}_{1}(j), Y^n_f(j-1)) \rightarrow X^n_1(j)$ which maps the message $\mathbf{M}_{1}(j)\in \mathcal{M}_{1}$ and feedback signal $Y'_{1}(j-1)$ to a codeword $X^n_1(j)\in \mathcal{X}^n_1$, 
		and $f_2$ at user 2, for each transmission: $\mathbf{M}_2(j) \rightarrow X^n_2(j)$, which maps the message $\mathbf{M}_2(j)\in \mathcal{M}_2$ to a codeword $X^n_2(j)\in \mathcal{X}^n_2(j)$.
		\item Two decoders, $g_1$ at user 1, for each transmission: $(Y^n_1, X^n_1) \rightarrow \hat{\mathcal{M}}_2$ which maps a received sequence $Y^n_1$ to a message $\hat{M}_2$;  $g_2$ at user 2: $(\mathcal{Y}^n_2, \mathcal{X}^n_2) \rightarrow \hat{\mathcal{M}}_{1}$ which maps a received sequence $Y^n_2$ to a message $\hat{M}_{1}$.
	\end{itemize}
			
	The two-way communication is governed by reliability and secrecy. The former is measured by the error probability and the secrecy levels, respectively. The {\em average error probabilities of decoding} at legitimate user 1 and 2 are defined as 
	\begin{align}
	P_{e,1}=&\frac{1}{2^{nR_{2}}} \sum_{\mathbf{M}_2=1}^{2^{nR_{2}}} Pr\{\hat{\mathbf{M}}_2\neq \mathbf{M}_2\}; \nonumber\\
	P_{e,2}=&\frac{1}{2^{n\overline{R}_{1s}}} \sum_{\mathbf{M}_{1}=1}^{2^{n\overline{R}_{1s}}} Pr\{\hat{\mathbf{M}}_1\neq \mathbf{M}_{1}\}. \label{Def_Error_Pr}
	\end{align}	
	
	The one-sided secrecy of the whole $b$ blocks is defined as
	\begin{align}
	\frac{1}{n}I((\mathbf{M}_1)_2^b;\mathbf{Z}^b)\leq \tau_n, \quad \lim\limits_{n\to\infty} \tau_n=0, \label{Def_One-sided_Secrecy_bBlock}
	\end{align} 
	where $(\mathbf{M}_1)_2^b=\{\mathbf{M}_{1}(2),\mathbf{M}_{1}(3),\cdots,\mathbf{M}_{1}(b)\}$ is the confidential messages from the $2$nd to the $b$-th block; $\mathbf{M}_1(j)$ denotes the $n$ length message in the $j$-th block; $\mathbf{Z}^b$ is the received messages at the eavesdropper in $b$ blocks.
	
	\begin{Remark}
		Note that the joint secrecy \cite{el2013achievable,pierrot2011strongly} is defined in one block as  
		\begin{equation}\label{Def_joint Secrecy}
		\frac{1}{n} I(\mathbf{M}_{1}, \mathbf{M}_2; \mathbf{Z})  \leq \tau_n, \quad \lim\limits_{n\to\infty} \tau_n=0, 
		\end{equation}
		where $\mathbf{M}_{1}$ and $\mathbf{M}_2$ are the message in one transmission of user 1 and user 2, respectively.
	\end{Remark}
	
	Comparing the one-sided secrecy constraint  \eqref{Def_One-sided_Secrecy_bBlock} with the joint secrecy constraint  \eqref{Def_joint Secrecy}, the joint secrecy requires the information leakage rate of both the messages $\mathbf{M}_{1}$ and $\mathbf{M}_2$ is demanded vanishing, while the one-sided secrecy requires the information leakage rate of the messages $\mathbf{M}_1)_2^b$ over $b-1$ blocks vanishing.  
		
	\begin{Definition}\label{Def_Achievable}
		The rate pair $(\overline{R}_{1s}, R_2)$ is said to be {\em achievable} under one-sided secrecy, if there exists a sequence of $(2^{n\overline{R}_{1s}},2^{nR_2},n)$ codes with $\overline{R}_{1s}=\frac{1}{n}H(\mathbf{M}_{1})$ and $R_2=\frac{1}{n}H(\mathbf{M}_2)$ such that
		\begin{align}		
		& P_{e,i} \leq \epsilon_n, \quad\mbox{for}\quad i=1,2 \label{Eqn_def_reliability}\\
		&\frac{1}{n}I((\mathbf{M}_{1})_2^b;\mathbf{Z}^b)  \leq \tau_n, \label{Eqn_def_one-sided secrecy}\\
		& \lim\limits_{n\to\infty} \epsilon_n=0 \quad \mbox{and}\quad \lim\limits_{n\to\infty} \tau_n=0. \label{Eqn_def limits}
		\end{align}
	\end{Definition}

\section{Main Result}\label{Sec_Main Result}	
In this section, we establish an achievable one-sided secrecy rate region for the two-way wiretap channel with feedback through multiple transmissions.	
An outer bound on secrecy capacity is also derived.

\subsection{One-sided security with cooperative coding}\label{Sec_Cooperative}

In order to bound the leakage rate through multiple transmissions, we first give the following lemma. 
\begin{Lemma}\label{Lemma_R1R2}
	Let $(U_1, U_2, Z)\sim p(u_1,u_2,z)$, and $\epsilon>0$. Denote $U_1^n(l_1),\ U_2^n(l_2)$, $l_1\in [1: 2^{n\overline{R}_1}],\ l_2 \in [1: 2^{n\overline{R}_2}]$ as random sequences of two users, and each distributed according to $\prod_{i=1}^n P_{U_1}(u_{1i}), \ \prod_{i=1}^n P_{U_2}(u_{2i})$. $\mathcal{C}_1,\ \mathcal{C}_2$ are the codebooks which contain the all $U_1^n(l_1),\ U_2^n(l_2)$ sequences, $\mathcal{C}_1=\{U_1^n(1),U_1^n(2),\ldots U_1^n(2^{n\overline{R}_1})\}$, $\mathcal{C}_2=\{U_2^n(1),U_2^n(2),\ldots U_2^n(2^{n\overline{R}_2})\}$. 	
	Let $L_1, L_2$ be random indexes with an arbitrary probability mass function. 	
	If $\overline{R}_1+\overline{R}_2 \geq I(U_1U_2;Z)$, $\overline{R}_1 \geq I(U_1;Z)$, $\overline{R}_2 \geq I(U_2;Z)$, then $H(L_1L_2|\mathcal{C}_1\mathcal{C}_2\mathbf{Z}(j)) \leq n(\overline{R}_1+\overline{R}_2-I(U_1U_2;Z)+\delta'(\varepsilon))$.
\end{Lemma}
\begin{IEEEproof}
	See Appendix \ref{Proof_Lemma_R1R2}.
\end{IEEEproof}	

Based on the lemma \ref{Lemma_R1R2}, considering the one-sided information leakage over $b$ blocks, i.e.  \eqref{Def_One-sided_Secrecy_bBlock}, we establish the following achievable secrecy rate region for the two-way wiretap channel with feedback through multiple transmissions. 

\begin{Theorem}\label{Thm_Inner_Coop}
	For the two-way wiretap channel with feedback under one-sided secrecy constraint, an achievable secrecy rate region is given by	
	\begin{align*}
	\mathcal{R}^{In}\stackrel{\vartriangle}{=}\mathbf{Conv} \{ \bigcup_{p\in \mathcal{P}} \mathcal{R}^{In}(p)\}
	\end{align*}
	where $\mathcal{P}$ denotes the set of all distribution of the random variables $U_1$, $U_2$, $X_1$, $X_2$ satisfying
	\begin{align*}
	p(q,u_1,x_1,u_2,x_2)=p(q)p(u_1|q)p(u_2|q)p(x_1|u_1)p(x_2|u_2).
	\end{align*} 
	$\mathcal{R}^{In}(p)$ is the region of rate pairs $(\overline{R}_{1s}, R_{2})$ for $p\in \mathcal{P}$, satisfying
	\begin{equation}\label{Equ: Inner_Individual}
	\left\{		
	\begin{aligned}
	&(\overline{R}_{1s}, R_{2}): \\
	&\overline{R}_{1s} \geq 0, R_{2} \geq 0,  \\
	&\overline{R}_{1s}\leq \min\{I(U_1;Y_2|X_2,Q)+I(U_2;Y_1|X_1,Q)-I(U_1U_2;Z),\quad I(U_1;Y_2|X_2,Q)-I(U_1;Z)\}, \\
	&R_2\leq I(U_2;Y_1|X_1) .\\
	&I(U_2;Y_1|X_1,Q)\geq I(U_2;Z).
	\end{aligned}
	\right\}	
	\end{equation}	
\end{Theorem}

\begin{IEEEproof}
	See the proof in Appendix \ref{Sec_Proof of Theorem 2}.
\end{IEEEproof}

Here,  we outline the proof to illustrate the main ideas. Our coding scheme involves the transmission of $b-1$ independent messages over $b$ $n$-transmission blocks.
\begin{itemize}
	\item During the $b$ blocks, user 1 sends a confidential message $\mathbf{M_1}(j)$ from the block $j=2$ to the block $j=b$. For each transmission, the message $\mathbf{M_1}(j)$ in the block $j\in [2:b]$ is split into two independent message pieces $(\mathbf{M}_{1u}(j), \mathbf{M}_{1s}(j))$, where $\mathbf{M}_{1u}(j)\in [1:2^{nR_{1u}}]$, $\mathbf{M}_{1s}(j)\in [1:2^{nR_{1s}}]$ with $\overline{R}_{1s}=R_{1u}+R_{1s}$. $R_{1u},\ R_{1s}$ are the transmission rates of $\mathbf{M}_{1u}(j)$ and $\mathbf{M}_{1s}(j)$, respectively, and $\overline{R}_{1s}$ denotes the secure transmission rate of user 1.  
	
	User 2 sends a confidential key $\mathbf{K}_2(1)$ in block $1$, a message $\mathbf{M}_{2}(j)$ and a confidential key  $\mathbf{K}_2(j)$ in block $2\leq j\leq b$ with $\mathbf{M}_{2}(j)\in [1:2^{nR_{2}}]$, $\mathbf{K}(j)\in [1:2^{nR_{2k}}]$ with $\overline{R}_{2}=R_{2}+R_{2k}+R_{2x}$. $R_{2},\ R_{2k}$ are the  transmission rates of $\mathbf{M}_{2}(j)$ and $\mathbf{K}(j)$, respectively.  
	
	\item The codebook of user 1 consists $2^{n\overline{R}_1}$ randomly generated i.i.d. sequences $u_1^{n}(l_1)$, $l_1\in [1:2^{n\overline{R}_1}]$. It is partitioned into $2^{nR_{1u}}$ equal-size bin $C_1(\mathbf{M}_{1u})$, further each bin is partitioned into $2^{nR_{1s}}$ equal-size sub-bin $C_1(\mathbf{M}_{1u},\mathbf{M}_{1s})$. 
	The codebook of user 2 consists $2^{n\overline{R}_2}$ randomly generated i.i.d. sequences $u_2^{n}(l_2)$, $l_2\in [1:2^{n\overline{R}_2}]$. It is equally partitioned into $2^{nR_{2}}$ bin $C_2(\mathbf{M}_{2})$, further each bin is equally partitioned into $2^{nR_{2k}}$ sub-bin $C_2(\mathbf{M}_{2},\mathbf{K}_2)$.
		
	\item In the first block, user 2  randomly selects a codeword $u_2^{n}(l_2)$ from the sub-bin $C_2(\mathbf{M_2}(1),\mathbf{K}_2(1))$, and generates the channel input $x_2^n(1)\sim \prod_{i=1}^n p(x_{21}(i)|u_2(i))$. User 1 does not send any message in the first block, only decodes the key $\mathbf{K}_2(1)$ at the end of the first block. Such that in the first block $R_{1u}=R_{1s}=R_{2}=0$. However, the impact on the whole achievable rate diminishes as the number of blocks $b\rightarrow \infty$.
	
	\item	In block $j\in [2:b]$, to send the message $\mathbf{M}_1(j)=(\mathbf{M}_{1u}(j), \mathbf{M}_{1s}(j))$, $\mathbf{M}_{1u}(j)$ is encrypted into $\mathbf{M}'_{1u}(j)$ by $\mathbf{K}_2(j-1)$ received from the previous block from user 2, i.e. $\mathbf{M}'_{1u}(j)=\mathbf{M}_{1u}(j)\oplus \mathbf{K}_2(j-1)$. A codeword $u_1^{n}(l_1)$ in block $j$ is randomly selected in sub-bin $C_1(\mathbf{M}'_{1u}(j), \mathbf{M}_{1s}(j))$. The channel input $x_1^n(j)$ is generated by $x_1^n(j)\sim \prod_{i=1}^n p(x_{1j}(i)|u_1(i))$.  
	On the other hand, user 2 sends a key $\mathbf{K}_2(j)$ in addition to the public message $\mathbf{M_2}(j)$. The codeword $u_2^{n}(l_2)$ is randomly chosen in sub-bin $C_2(\mathbf{M_2}(j),\mathbf{K}_2(j))$. Then the channel input $x_2^n(j)$ is generated by $x_2^n(j)\sim \prod_{i=1}^n p(x_{2j}(i)|u_2(i))$.
	
	In the coding scheme, the cooperative jamming is used to improve the secrecy rate of the confidential message of user 1 by the jamming from user 2. 
	The cooperative jamming can be interpreted by channel prefixing mechanism \cite{ulukus2009cooperative}, which means prefixing an artificial discrete memoryless channel before the communication channel. 
	In the proposed coding scheme, the codewords $u_1^{n}$ and $u_2^{n}$ are drawn from two random binning codebooks, and passed into two prefixing channels to generate the channel inputs $x_1^n$ and $x_2^n$, respectively. 
	
	\item Each decoder uses the joint typical decoding together with the knowledge of its own codewords. 
	In each block, the legitimate user decodes the channel output and obtains the message from the other user with rather small average error probability.
	\item From block 2 to block $b$, $b-1$ confidential messages $(\mathbf{M}_1)_{2}^b$ are sent and should be kept secure from the eavesdropper. The one-sided secrecy constraint through the whole $b$ block is measured by 
	\begin{align}\label{Inf_Leak_n}
	\frac{1}{n}I((\mathbf{M}_1)_{2}^b;\mathbf{Z}^b|\mathcal{C}_1\mathcal{C}_2) \leq \tau_n,  \quad \lim\limits_{n\to\infty} \tau_n=0.
	\end{align}
\end{itemize}

Applying theorem \ref{Thm_Inner_Coop} to a special two-way channel where the legitimate users and eavesdropper have the same channel output, i.e. $Y_1=Y_2=Z$, we have the following corollary. 

\begin{Corollary}\label{Cor_InnerCo}
	For the two-way wiretap channel, if the legitimate users and eavesdropper have the same channel output, i.e. $Y_1=Y_2=Z$, an achievable one-sided secrecy rate region is the union of non-negative rate pairs $(\overline{R}_{1s}, R_2)\in \mathcal{R}_1^{In}$ satisfying
	\begin{align*}
	R_2 \leq & I(X_2; Z|X_1) \\
	\overline{R}_{1s} \leq & I(U_1;Z|X_2)+I(U_2;Z|X_1)-I(U_1U_2;Z) \quad \text{if} ~~ I(U_2;Z|X_1)< I(U_2;Z|U_1) \\
									& I(U_1;Z|X_2)-I(U_1;Z)  \quad \text{if} ~~ I(U_2;Z|X_1)\geq I(U_2;Z|U_1) \\
	\end{align*}
	over all $p(x_1,x_2)$.
\end{Corollary}


\subsection{An outer bound of two-way wiretap channel}\label{Sec_OuterBound}
	\begin{Theorem}\label{Thm_OuterBound}
		For the two-way wiretap channel with feedback under one-sided secrecy constraint, an outer bound on secrecy capacity region is  
		\begin{align*}
		\mathcal{R}^O\stackrel{\vartriangle}{=}\mathbf{Conv} \{ \bigcup_{p\in \mathcal{P}} \mathcal{R}^O(p)\}
		\end{align*}	
		where $\mathcal{P}$ denotes the set of all distribution of the random variables $U$, $V$, $X_1$, $X_2$ satisfying
		\begin{align*}
		p(q,u,v,x_1,x_2)=p(q)p(u|q)p(v|u)p(x_1x_2|uv).
		\end{align*} 
		$\mathcal{R}^O(p)$ is the region of rate pairs $(\overline{R}_{1s}, R_2)$ for $p\in \mathcal{P}$, satisfying 
		\begin{align}
		R_2 \leq &I(X_2;Y_1|X_1), \nonumber\\
		\overline{R}_{1s} \leq &\min \{ I(V; X_{2}, Y_{2}|U)-I(V;Z|U), \quad I(V; X_{1}, Y_{1}|U)-I(V;Z|U)\},  \label{Equ_Conv_R1e}
		\end{align}
		where $U\to V\to (X_1, X_2)\to (Y_1, Y_2, Z)$ forms a Markov Chain.
	\end{Theorem}
	\begin{IEEEproof}
		See the proof in Appendix \ref{Proof_Outer}.
	\end{IEEEproof}
	
	\begin{Corollary}\label{Cor_Outer_Y=Z}
		For the two-way wiretap channel, if the legitimate users and eavesdropper have the same channel output, i.e. $Y_1=Y_2=Z$, the outer bound on secrecy capacity region is the union of non-negative rate pairs $(\overline{R}_{1s}, R_2)\in \mathcal{R}^{O}_1$ satisfying  
		\begin{align*}
		R_2 \leq &I(X_2;Z|X_1),  \\
		\overline{R}_{1s}\leq &\min\{H(X_1|Z), H(X_2|Z)\}   
		\end{align*}  
	\end{Corollary}
	\begin{IEEEproof} Applying the outer bound given in \eqref{Equ_Conv_R1e}, we have
		\begin{align*}
		\overline{R}_{1s}\stackrel{(a)}=& I(V; X_{2}, Z| U)-I(V;Z|U)\\
		= & I(V;X_2|U,Z) \\
		\leq & H(X_2|U,Z) \\
		\stackrel{(b)}\leq & H(X_2|Z)
		\end{align*}
		where $(a)$ follows from $Z=Y_2$; $(b)$ is due to the fact that conditioning does not increase entropy.
		A similar proof can be applied to show that $\overline{R}_{1s}\leq H(X_1|Z)$.
	\end{IEEEproof}
	
\section{Numerical examples}\label{Sec: BiCase}
In this section, we consider the binary-input two-way wiretap channels for numerical illustrations. In particular, we assume that all the terminals have the same observations, i.e., $Y_1=Y_2=Z$.

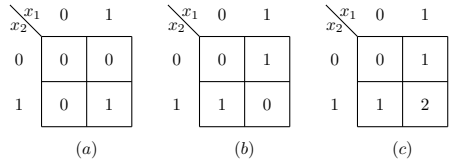
\begin{figure}[htbp]
	\centering
	\begin{tikzpicture}[node distance=2cm,auto,>=latex', scale=0.6, every node/.style={scale=0.65}]
	\draw (0.3,-0.3) -- (1,-1);
	\draw (1,-1) -- (3,-1);
	\draw (1,-1) -- (1,-3);
	\draw (3,-1) -- (3,-3);
	\draw (1,-3) -- (3,-3);
	\draw (1,-2) -- (3,-2);
	\draw (2,-1) -- (2,-3);
	\node at (0.8,-0.5) {$x_1$};
	\node at (0.5,-0.8) {$x_2$};
	\node at (1.5,-0.5) {$0$};
	\node at (2.5,-0.5) {$1$};
	\node at (0.5,-1.5) {$0$};
	\node at (0.5,-2.5) {$1$};
	
	\node at (1.5,-1.5) {$0$};
	\node at (2.5,-1.5) {$0$};
	\node at (1.5,-2.5) {$0$};
	\node at (2.5,-2.5) {$1$};
	\node at (2,-3.5) {$(a)$};
	
	\draw (3.8,-0.3) -- (4.5,-1);
	\draw (4.5,-1) -- (6.5,-1);
	\draw (4.5,-1) -- (4.5,-3);
	\draw (6.5,-1) -- (6.5,-3);
	\draw (4.5,-3) -- (6.5,-3);
	\draw (4.5,-2) -- (6.5,-2);
	\draw (5.5,-1) -- (5.5,-3);
	\node at (4.3,-0.5) {$x_1$};
	\node at (4,-0.8) {$x_2$};
	\node at (5,-0.5) {$0$};
	\node at (6,-0.5) {$1$};
	\node at (4,-1.5) {$0$};
	\node at (4,-2.5) {$1$};
	
	\node at (5,-1.5) {$0$};
	\node at (6,-1.5) {$1$};
	\node at (5,-2.5) {$1$};
	\node at (6,-2.5) {$0$};
	\node at (5.5,-3.5) {$(b)$};
	
	\draw (7.3,-0.3) -- (8,-1);
	\draw (8,-1) -- (10,-1);
	\draw (8,-1) -- (8,-3);
	\draw (10,-1) -- (10,-3);
	\draw (8,-3) -- (10,-3);
	\draw (8,-2) -- (10,-2);
	\draw (9,-1) -- (9,-3);
	\node at (7.8,-0.5) {$x_1$};
	\node at (7.5,-0.8) {$x_2$};
	\node at (8.5,-0.5) {$0$};
	\node at (9.5,-0.5) {$1$};
	\node at (7.5,-1.5) {$0$};
	\node at (7.5,-2.5) {$1$};
	
	\node at (8.5,-1.5) {$0$};
	\node at (9.5,-1.5) {$1$};
	\node at (8.5,-2.5) {$1$};
	\node at (9.5,-2.5) {$2$};
	\node at (9,-3.5) {$(c)$};
	
	\end{tikzpicture}
	\vspace{-2mm}
	\caption{Transition diagram of the binary-input two-way channels.}
	\label{Fig_BiCase}
	\vspace{-3mm}
\end{figure}

If restricting to the binary-input (i.e., $x_1, x_2\in \{0,1\}$) and binary-output (i.e., $y_1=y_2=z\in \{0,1\}$) deterministic two-way channels, there are in total 16 transition possibilities. Among them, only two transition possibilities could have positive transmission rates at both legitimate users. The transmission diagrams of these channels are shown in Fig. \ref{Fig_BiCase} $(a)$ and $(b)$, referred as the binary Multiplying channel (BMC) and the XOR channel, respectively.
If allowing ternary outputs (i.e., $y_1=y_2=z\in \{0,1,2\}$), the Adder channel, as shown in  Fig. \ref{Fig_BiCase} $(c)$,
is also interesting.

For each channel, the numerical result consists of the achievable secrecy rate region with feedback (according to Theorem \ref{Thm_Inner_Coop}), the achievable secrecy rate region without feedback and the outer bound on secrecy capacity (according to Corollary \ref{Cor_Outer_Y=Z}).

\subsection{Binary Multiplying channel}
The transmission diagrams of BMC is shown in Fig. \ref{Fig_BiCase} (a), where the channel output is represented by $Y_1=Y_2=Z=X_1\cdot X_2$. 

By Theorem \ref{Thm_Inner_Coop}, the achievable secrecy rate region $\mathcal{R}^{Co-in}$ with feedback is drawn with $X_1X_2\sim \Pr (p_1p_2)$. Correspondingly the outer bound $\mathcal{R}^{O}$ on secrecy capacity and the achievable secrecy rate region $\mathcal{R}^{Ind-in}$ without feedback are also derived for comparison. 

\begin{figure}[!h]
	\centering
	\includegraphics[width=0.5\textwidth]{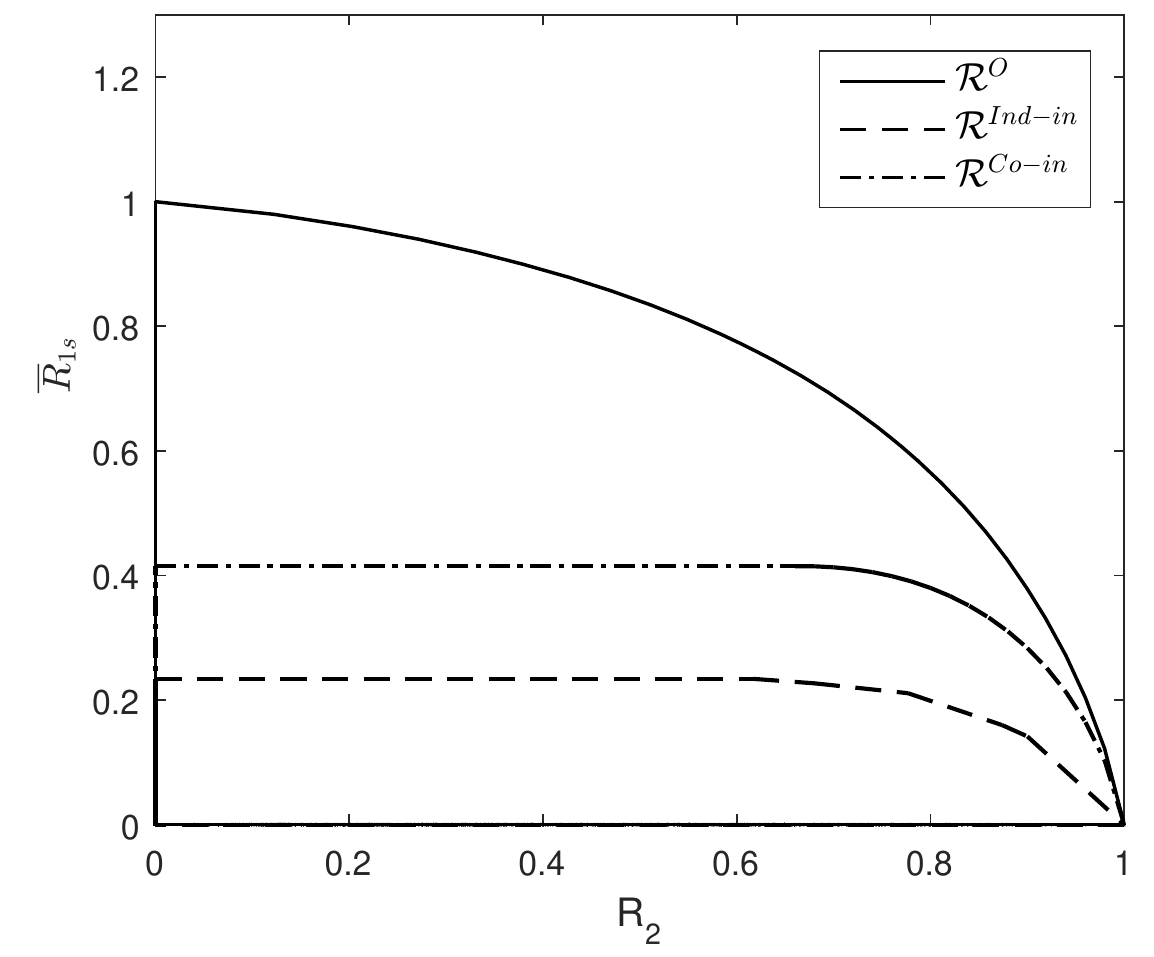}
	\caption{BMC with one-sided secrecy: achievable regions \& outer bound.}
	\label{fig: Multiple}
\end{figure}

Seen from Fig. \ref{fig: Multiple}, clearly the achievable rate region $\mathcal{R}^{Co-in}$ is larger than $\mathcal{R}^{Ind-in}$, which indicates the benefit of feedback in improving the secrecy rate in two-way wiretap channel. This phenomenon is more obviously for the binary XOR channel. However the gap between the achievable rate region $\mathcal{R}^{Co-in}$ and the outer bound $\mathcal{R}^{O}$ is still large.  

\subsection{Binary XOR channel}
The channel is shown in Fig. \ref{Fig_BiCase} (b), where the channel output is represented by $Y_1=Y_2=Z=X_1\oplus X_2$. For the Binary XOR channel, we draw the curves of the achievable secrecy regions $\mathcal{R}^{Ind-in}$ without feedback, $\mathcal{R}^{Co-in}$ with feedback and the outer bound $\mathcal{R}^{O}$, respectively. 
 
\begin{figure}[!htbp]
	\centering
	\includegraphics[width=0.5\textwidth]{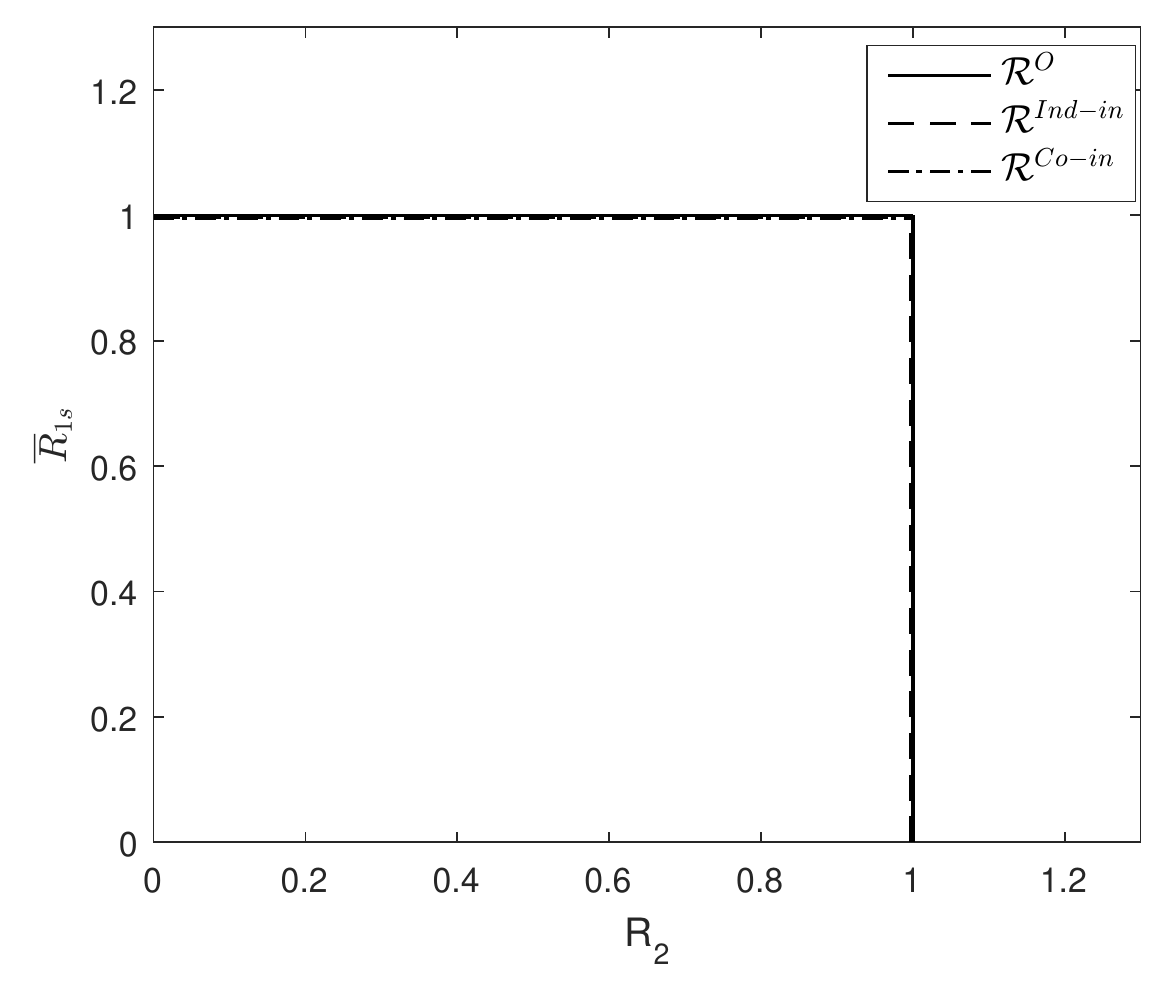}
	\caption{Binary XOR channel with one-sided secrecy: achievable regions \& outer bound.}
	\label{fig: XOR}
\end{figure}

In Fig. \ref{fig: XOR}, both $\mathcal{R}^{Co-in}$ and $\mathcal{R}^{Ind-in}$ can achieve the maximum transmission rate i.e., $\max \overline{R}_{1s}= \max R_2=1$. 
Specially, the point $(\overline{R}_{1s}, R_2)=(1, 0)$ can be achieved by $C_2\sim \mbox{Bern}(1),$  $X_2\sim \mbox{Bern}(1/2)$ and $C_1=X_1\sim \mbox{Bern}(1/2)$. At this point, user 2 is transmitting random bits (i.e., $X_2\sim \mbox{Bern}(1/2)$) but messages (since $C_2\sim \mbox{Bern}(1)$ and thus $R_2=0$). These random bits work as the cooperative jamming to help the secret transmission of user 1. In this way, the secret transmission rate $\overline{R}_{1s}$ is increased. 
Note that, according to Corollary \ref{Cor_Outer_Y=Z}, the outer bound satisfies $\overline{R}_{1s}\leq 1, R_2\leq 1$, coinciding with $\mathcal{R}^{Co-in}$. Therefore, for the Binary XOR channel, the one-sided secrecy capacity region is established. 

\subsection{Adder channel}
\begin{figure}[!htbp]
	\centering
	\includegraphics[width=0.5\textwidth]{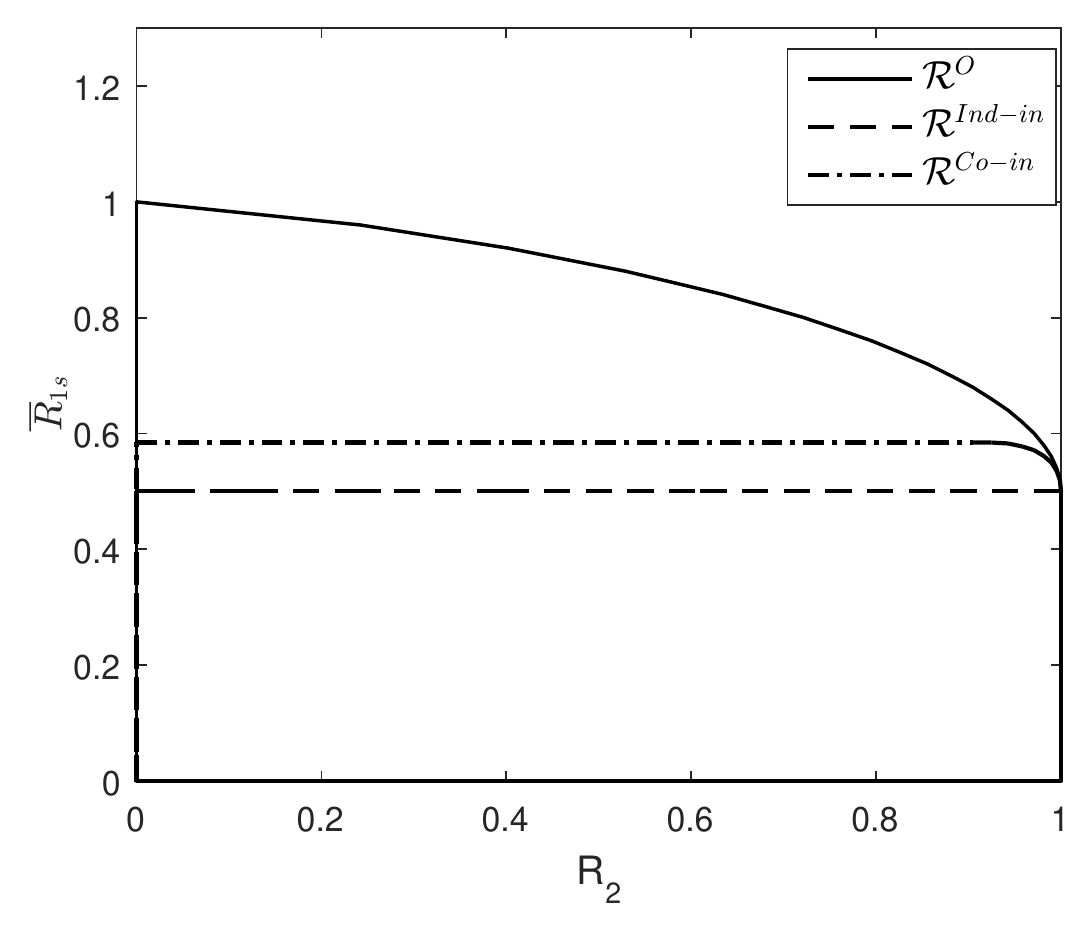}
	\caption{Adder channel with one-sided secrecy: achievable regions \& outer bound.}
	\label{fig: ADD}
\end{figure}
The channel is shown in Fig. \ref{Fig_BiCase} (c), where the channel output is represented by $Y_1=Y_2=Z=X_1+X_2$.
The achievable one-sided secrecy rate regions $\mathcal{R}^{Ind-in}$ without feedback, $\mathcal{R}^{Co-in}$ with feedback and the outer bound $\mathcal{R}^{O}$ are shown in Fig. \ref{fig: ADD}.

Similar to the behavior at the BMC, the achievable rate region $\mathcal{R}^{Co-in}$ with feedback is larger than $\mathcal{R}^{Ind-in}$ without feedback. 
Under the one-sided secrecy constraint, the maximum achievable rates at both users are $\max \overline{R}_{1s} =0.5$ and $\max R_2=1,$ respectively. Remarkably, for user 1, there is $50\%$ rate loss compared to the maximum transmission rate in case of no secrecy constraint.

\appendices

\section{Proof of Lemma \ref{Lemma_R1R2} } \label{Proof_Lemma_R1R2}
\begin{IEEEproof} For the simultaneous joint typical decoding, we first define the following  events:
	\begin{itemize}
		\item $\mathcal{E}=\{(U_1^n(L_1), U_2^n(L_2), Z^n)\notin T_\epsilon^{(n)} \}$;
		\item $\mathcal{E}_1=\{(U_1^n(L_1), U_2^n(L_2), Z^n)\in T_\epsilon^{(n)} \}$;
		\item $\mathcal{E}_2=\{(U_1^n(t_1), U_2^n(t_2), Z^n)\in T_\epsilon^{(n)}, t_1\neq L_1, t_2\neq L_2 \}$;
		\item $\mathcal{E}_3=\{(U_1^n(L_1), U_2^n(t_2), Z^n)\in T_\epsilon^{(n)}, t_2\neq L_2 \}$;
		\item $\mathcal{E}_4=\{(U_1^n(t_1), U_2^n(L_2), Z^n)\in T_\epsilon^{(n)}, t_1\neq L_1 \}$;	
	\end{itemize}
	
	Based on the definition of the events above, we have the following properties.
	
	\begin{itemize}
		\item By LLN, the first event $P\{E=1\} =P\{\mathcal{E}\}$ tends to zero as $n\rightarrow \infty$.
		
		\item For event $\mathcal{E}_1$, define 
		\begin{align*}
		N(Z^n, U_1^n(L_1), U_2^n(L_2))=&|\{(U_1^n(L_1), U_2^n(L_2), Z^n)\in T_\epsilon^{(n)}\}|=1.
		\end{align*}	
		
		\item For event $\mathcal{E}_2$, define 
		\begin{align*}
		N(Z^n, U_1^n(t_1), U_2^n(t_2))=&|\{t_1\in [1:2^{n\overline{R}_1}], t_2\in [1:2^{n\overline{R}_2}]: \\
		& (U_1^n(t_1), U_2^n(t_2), Z^n)\in T_\epsilon^{(n)}, t_1\neq L_1, t_2\neq L_2\}|\\
		\leq & \sum_{t_1=1, t_1\neq L_1}^{2^{n\overline{R}_1}} \sum_{t_2=1, t_2\neq L_2}^{2^{n\overline{R}_2}} \Pr\{(U_1^n(t_1), U_2^n(t_2), Z^n)\in T_\epsilon^{(n)}\}\\	
		\leq & 2^{n(\overline{R}_1+\overline{R}_2-I(U_1U_2;Z)+\delta(\varepsilon))}.
		\end{align*}		
		
		\item For event $\mathcal{E}_3$, define 
		\begin{align*}
		N(Z^n, U_1^n(L_1), U_2^n(t_2))=&|\{t_2\in [1:2^{n\overline{R}_2}]:  (U_1^n(L_1), U_2^n(t_2), Z^n)\in T_\epsilon^{(n)}\}| \\
		\leq & \sum_{t_2=1, t_2\neq L_2}^{2^{n\overline{R}_2}} \Pr\{(U_1^n(L_1), U_2^n(t_2), Z^n)\in T_\epsilon^{(n)}\}\\
		\leq & 2^{n(\overline{R}_2-I(U_2;ZU_1)+\delta(\varepsilon))}.
		\end{align*}
		
		\item For event $\mathcal{E}_4$, define 
		\begin{align*}
		N(Z^n, U_1^n(t_1), U_2^n(L_2))=&|\{t_1\in [1:2^{n\overline{R}_1}]: (U_1^n(t_1), U_2^n(L_2), Z^n)\in T_\epsilon^{(n)}\}|\\
		\leq & \sum_{t_1=1, t_1\neq L_1}^{2^{n\overline{R}_1}} \Pr\{(U_1^n(t_1), U_2^n(L_2), Z^n)\in T_\epsilon^{(n)}\}\\
		\leq & 2^{n(\overline{R}_1-I(U_1;ZU_2)+\delta(\varepsilon))}
		\end{align*}
	\end{itemize}	

	Define $E=1$ if $\mathcal{E}$ occurs.
	$E=0$ if events $\mathcal{E}_1$ occures and any of $\mathcal{E}_2,\mathcal{E}_3,\mathcal{E}_4$ occur. 
	
	Hence,
	\begin{align}
	& P(E=0)H(L_1L_2|\mathcal{C}_1\mathcal{C}_2\mathbf{Z}(j),E=0)  \nonumber \\
	\leq & P(E=0) \log_2[1+2^{n(\overline{R}_1+\overline{R}_2-I(U_1U_2;Z)+\delta(\varepsilon))}+ 2^{n(\overline{R}_2-I(U_2;ZU_1)+\delta(\varepsilon))}+2^{n(\overline{R}_1-I(U_1;ZU_2)+\delta(\varepsilon))}]   \nonumber\\
	\leq &\log_2[2^{n(\overline{R}_1+\overline{R}_2-I(U_1U_2;Z)+\delta(\varepsilon))}(2^{-n(\overline{R}_1+\overline{R}_2-I(U_1U_2;Z)+\delta(\varepsilon))}+1+2^{-n(\overline{R}_2-I(U_2;Z))}+2^{-n(\overline{R}_1-I(U_1;Z))})]   \nonumber\\
	\stackrel{(a)} \leq &n[\overline{R}_1+\overline{R}_2-I(U_1U_2;Z)+\delta(\varepsilon)] +2 \label{Equ_Lemma_PE0H}
	\end{align}
	where (a) follows from that if $\overline{R}_1+\overline{R}_2\geq I(U_1U_2;Z)$, $\overline{R}_2\geq I(U_2;Z)$ and $\overline{R}_1\geq I(U_1;Z)$.
	
	Therefore,     
	\begin{align*}
	& H(L_1L_2E|\mathcal{C}_1\mathcal{C}_2\mathbf{Z}(j)) \\
	=& H(E|\mathcal{C}_1\mathcal{C}_2)+H(L_1L_2|\mathcal{C}_1\mathcal{C}_2\mathbf{Z}(j)E)\\
	\leq & 1+ P(E=1)H(L_1L_2|\mathcal{C}_1\mathcal{C}_2\mathbf{Z}(j),E=1)+ P(E=0)H(L_1L_2|\mathcal{C}_1\mathcal{C}_2\mathbf{Z}(j),E=0)\\
	\stackrel{(a)} \leq & 1+ P(E=1)n(\overline{R}_1+\overline{R}_2)+ n[\overline{R}_1+\overline{R}_2-I(U_1U_2;Z)+\delta(\varepsilon)] +2  \\
	\stackrel{(b)} =& 1+n(\overline{R}_1+\overline{R}_2-I(U_1U_2;Z)+\delta(\varepsilon)) +2   \\
	= & n[\overline{R}_1+\overline{R}_2-I(U_1U_2;Z)+\delta(\varepsilon)+\dfrac{1}{n}]+\delta'_1(\varepsilon)
	\end{align*}
	where (a) the second term follows from that $H(L_1L_2|\mathcal{C}_1\mathcal{C}_2\mathbf{Z}(j),E=1)\leq n(\overline{R}_1+\overline{R}_2)$, and the third term follows from the above  \eqref{Equ_Lemma_PE0H};
	(b) follows from that $P(E=1)\rightarrow 0$ as $n\rightarrow \infty$.
	Such that $H(L_1L_2|\mathcal{C}_1\mathcal{C}_2\mathbf{Z}(j)) \leq n(\overline{R}_1+\overline{R}_2-I(U_1U_2;Z)+\delta'(\varepsilon))$, if $\overline{R}_1+\overline{R}_2\geq I(U_1U_2;Z)$, $\overline{R}_2\geq I(U_2;Z)$ and $\overline{R}_1\geq I(U_1;Z)$.
	
\end{IEEEproof}

\section{Proof of Theorem \ref{Thm_Inner_Coop}}\label{Sec_Proof of Theorem 2} 
The transmission is performed for $B$ blocks of length $n$, where both $B$ and $n$ are sufficiently large.
With fixed probabilities density function $p(q)$, $p(u_1|q)$ and $p(u_2|q)$, the random code generation is described as follows.

\subsection{Codebook Generation:}

\subsubsection{Codebook generation}
	With the fixed $p(q)$, generate a sequence $q^{n}$,  where each of its element is i.i.d. and randomly chosen according to $p(q)$. The sequence $q^{n}$ is then sent to two users before the communication.

\begin{itemize}
	\item User 1: For a given distribution $p(u_1|q)$ and the sequence $q^{n}$, randomly generate i.i.d. sequences $u_1^{n}(l_1)=u_1^{n}(\mathbf{M}_{1u},\mathbf{M}_{1s},\mathbf{M}_{1x})$ where $l_1\in [1:2^{n\overline{R}_1}]$, $\overline{R}_1\geq R_{1u}+R_{1s}$, with $\overline{R}_1= R_{1u}+R_{1s}+R_{1x}$.
	Partition sequences $u_1^{n}(l_1)$ ($l\in [1:2^{n\overline{R}_1}]$) into $2^{nR_{1u}}$ equal-size bin $C_1(\mathbf{M}_{1u})$, where $\mathbf{M}_{1u}\in [1:2^{nR_{1u}}]$.
	Further partition each bin $C_1(\mathbf{M}_{1u})$ into $2^{nR_{1s}}$ equal-size sub-bin $C_1(\mathbf{M}_{1u},\mathbf{M}_{1s})$.	
	
	\item User 2:  For a given distribution $p(u_2|q)$ and the sequence $q^{n}$, generate i.i.d. sequences $u_2^{n}(l_2)=u_2^{n}(\mathbf{M}_{2},\mathbf{K}_2,\mathbf{M}_{2x})$ where $l_2\in [1:2^{n\overline{R}_2}]$, $\overline{R}_2\geq R_{2}+R_{2k}$, let $\overline{R}_2=R_{2}+R_{2k}+R_{2x}$.
	Partition sequences $u_2^{n}(l_2)$ ($[1:2^{n\overline{R}_2}]$) into $2^{nR_{2}}$ equal-size bin $C_2(\mathbf{M}_{2})$, where $\mathbf{M}_{2}\in [1:2^{nR_{2}}]$.
	Further partition each bin $C_2(\mathbf{M}_{2})$ into $2^{nR_{2k}}$ equal-size sub-bin $C_2(\mathbf{M}_{2},\mathbf{K}_2)$.	
\end{itemize}

\subsection{Encoding}
We use block coding scheme to transmit messages over $b$ transmissions blocks. 	
In the first block, only user 2 sends a key message $\mathbf{K}_2(1)$ to user 1. The user 2 randomly selects an index $l_2 \in C_2(\mathbf{M}_2(1),\mathbf{K}_2(1))$; and uses the corresponding $u_2^{n}(l_2)$ to generate the symbol $x_2^n(1)\sim \prod_{i=1}^n p(x_{2i}|u_{2i})$. And send the symbols $x_2^n(1)$ to user 1.

In the next block $j\geq 2$, the encoding scheme is described as follows. 
Suppose that user 1 intends to send the confidential message $\mathbf{M_1}(j)=(\mathbf{M}_{1u}(j), \mathbf{M}_{1s}(j))$. Then it encodes the message in the following steps.
\begin{enumerate}
	\item  $\mathbf{M}_{1u}(j)$ is encrypted into $\mathbf{M}'_{1u}(j)$ by the key $\mathbf{K}_2(j-1)$ received from user 2 in block $j-1$, as $\mathbf{M}'_{1u}(j)=\mathbf{M}_{1u}(j)\oplus \mathbf{K}_2(j-1)$.			
	Under security constraint, it must satisfy $R_{1u}\leq R_{2k}$.
	\item A codeword $u_1^{n}(l_1)$ for block $j$ is randomly selected from sub-bin $C_1(\mathbf{M}'_{1u}(j), \mathbf{M}_{1s}(j))$. The channel input $x_1^n(j)$ is generated by $x_1^n(j)\sim \prod_{i=1}^n p(x_{1i}|u_{1i})$.
\end{enumerate}

Suppose that user 2 intends to send the message $\mathbf{M}_2(j)$, and the key $\mathbf{K}_2(j) \in \mathcal{K}$. The codeword $u_2^{n}(l_2)$ is randomly chosen from sub-bin $C_2(\mathbf{M_2}(j),\mathbf{K}_2(j))$. Then the channel input $x_2^n(j)$ is generated by $x_2^n(j)\sim \prod_{i=1}^n p(x_{2i}|u_{2i})$.

\subsection{Decoding}
In the $j$-th block, user 1 declares that $\hat{\mathbf{M}}_2(j)$ is sent by user 2 if $u_2^{n}(\hat{\mathbf{M}}_2(j),\mathbf{K}_2(j), \mathbf{M}_{2x}(j))$ is the unique sequence  such that $(q^{n},u_2^{n}(\hat{\mathbf{M}}_2(j),\mathbf{K}_2(j), \mathbf{M}_{2x}(j)), y_1^{n}(j), x_1^n(j))\in T^{n}_{1,\varepsilon}$. 
User 2 declares that $\hat{\mathbf{M}}'_{1u}(j)$ is sent by user 1 if $u_1^{n}(\hat{\mathbf{M}}'_{1u}(j), \mathbf{M}_{1s}(j), \mathbf{M}_{1x}(j))$ is the unique sequence such that $(q^{n},u_1^{n}(\hat{\mathbf{M}}'_{1u}(j), \mathbf{M}_{1s}(j), \mathbf{M}_{1x}(j)), y_2^{n}(j), x_2^n(j))\in T^{n}_{2,\varepsilon}$. User 2 recovers $\mathbf{M}_{1u}(j)$ by $\mathbf{M}_{1u}(j)=\hat{\mathbf{M}}'_{1u}(j)\oplus \mathbf{K}_2(j-1)$.

\subsection{Reliability}
Assume that $\mathbf{M}_{1s}(j)=1,\ \mathbf{M}'_{1u}(j)=1, \ \mathbf{M}_{2}(j)=1, \ \mathbf{K}_2(j)=1$ are sent, and the corresponding codeword $u_1^{n}(l_1=1), \ u_2^{n}(l_2=1)$.

First we consider the error probability at user 1. A decoding error happens at user 1 if at least one of the following events occur 
\begin{enumerate}
	\item $\mathcal{E}_{11}$: Given $q^{n}$, the codeword $u_2^{n}(1)$ and $(x_1^n, y_1^n)$ are not jointly typical, i.e. $\mathcal{E}_{11}=\{(q^{n},u_2^{n}(l_2), y_1^{n}, x_1^n)\notin T^{n}_{1,\varepsilon}\}$;
	\item $\mathcal{E}_{1i}, i\neq 1$: Given $q^{n}$, some other codewords $u_2^{n}(i)$ and $(x_1^n, y_1^n)$ are jointly typical, i.e., $\mathcal{E}_{1i}=\{(q^{n},u_2^{n}(i), x_1^n, y_1^n)\in T_\epsilon^n, i\neq 1\}$.
\end{enumerate}
Hence, $P(\mathcal{E}_1)$ at user 1 can be bounded by 
\begin{align*}
P(\mathcal{E}_1)  \leq & \Pr\{\mathcal{E}_{11}\} +\sum_{i=2}^{2^{n\overline{R}_2}}\Pr\{\mathcal{E}_{1i}|\mathcal{E}_{11}^C \} \\
\stackrel{(a)}\leq  & \epsilon + 2^{n[\overline{R}_2-I(U_2;Y_1|X_1,Q)+3\epsilon]} \\
\stackrel{(b)}\leq  & \delta_1(\epsilon),
\end{align*}
where $(a)$ follows from the LLN; and $(b)$ is by the packing lemma \cite{el2013achievable} if taking $\overline{R}_2\leq I(U_2;Y_1|X_1,Q)$.

Similarly. for user 2, the  error events are defined as follows
\begin{enumerate}
	\item $\mathcal{E}_{21}$: Given $q^{n}$, the codeword $u_1^{n}(1)$ and $(x_2^n, y_2^n)$ are not jointly typical, i.e. $\mathcal{E}_{21}=\{(q^{n},u_1^{n}(1), y_2^{n}, x_2^n)\notin T^{n}_{2,\varepsilon}\}$;
	\item $\mathcal{E}_{2i}, i\neq 1$: Given $q^{n}$, some other codewords $u_1^{n}(i)$ and $(x_2^n, y_2^n)$ are jointly typical, i.e., $\mathcal{E}_{2i}=\{(q^{n},u_1^{n}(i), x_2^n, y_2^n)\in T_\epsilon^n, i\neq 1\}$.
\end{enumerate}

By the union bound, the error probability $P(\mathcal{E}_2)$ at user 2 are
\begin{align*}
P(\mathcal{E}_2)  \leq & \Pr\{\mathcal{E}_{21}\} +\sum_{i=2}^{2^{n\overline{R}_1}}\Pr\{\mathcal{E}_{2i}|\mathcal{E}_{21}^C \} \\
\stackrel{(a)}\leq  & \epsilon + 2^{n[\overline{R}_1-I(U_1;Y_2|X_2,Q)+3\epsilon]} \\
\stackrel{(b)}\leq  & \delta_1(\epsilon),
\end{align*}
where $(a)$ follows from the LLN; and $(b)$ is by the packing lemma \cite{el2013achievable} if taking $\overline{R}_1\leq I(U_1;Y_2|X_2,Q)$.

\subsection{Information Leakage Rate}
Now we bound the information leakage rate through $b$ blocks by the block coding scheme. During the whole $b$ blocks, from block 2 to block $b$, $b-1$ confidential messages $(\mathbf{{M_1}})_{2}^b$ are sent and should be kept secure from the eavesdropper. Here we use $I((\mathbf{{M_1}})_{2}^b;\mathbf{Z}^b|\mathcal{C}_1\mathcal{C}_2)$ to denote the information leakage through the whole $b$ blocks, and the information leakage rate are considered averaged over the codes.

\begin{align*}
& I((\mathbf{{M_1}})_{2}^b;\mathbf{Z}^b|\mathcal{C}_1\mathcal{C}_2) \\
=& \sum_{j=2}^b I(\mathbf{M_1}(j);\mathbf{Z}^b|\mathcal{C}_1\mathcal{C}_2(\mathbf{M_1})_{j+1}^b) \\
\stackrel{(a)} \leq & \sum_{j=2}^b I(\mathbf{M_1}(j);\mathbf{Z}^b|\mathcal{C}_1\mathcal{C}_2(\mathbf{M_1})_{j+1}^b\mathbf{K}_2(j)) \\
=& \sum_{j=2}^b I(\mathbf{M_1}(j);\mathbf{Z}^j|\mathcal{C}_1\mathcal{C}_2(\mathbf{M_1})_{j+1}^b\mathbf{K}_2(j))+I(\mathbf{M_1}(j);\mathbf{Z}_{j+1}^b|\mathcal{C}_1\mathcal{C}_2(\mathbf{M_1})_{j+1}^b\mathbf{K}_2(j)\mathbf{Z}^j) \\
\stackrel{(b)} =& \sum_{j=2}^b I(\mathbf{M_1}(j);\mathbf{Z}^j|\mathcal{C}_1\mathcal{C}_2(\mathbf{M_1})_{j+1}^b\mathbf{K}_2(j)) \\
\stackrel{(c)} =& \sum_{j=2}^b I(\mathbf{M_1}(j);\mathbf{Z}^j|\mathcal{C}_1\mathcal{C}_2\mathbf{K}_2(j))
\end{align*}
where $(a)$ follows that $\mathbf{M_1}(j)$ is independent with $(\mathbf{K}_2(j), (\mathbf{M_1})_{j+1}^b)$;   
$(b)$ follows by Markov chain $(\mathbf{Z}_{j+1}^b,(\mathbf{M_1})_{j+1}^b,\mathcal{C}_1,\mathcal{C}_2)\rightarrow (\mathbf{Z}^j,\mathbf{K}_2(j),\mathcal{C}_1,\mathcal{C}_2) \rightarrow (\mathbf{M_1}(j),\mathcal{C}_1,\mathcal{C}_2)$. 
$(c)$ follows that
\begin{align*}
& I(\mathbf{M_1}(j);\mathbf{Z}^j|\mathcal{C}_1\mathcal{C}_2(\mathbf{M_1})_{j+1}^b\mathbf{K}_2(j)) \\
= & H(\mathbf{M_1}(j)|\mathcal{C}_1\mathcal{C}_2\mathbf{K}_2(j)(\mathbf{M_1})_{j+1}^b)-H(\mathbf{M_1}(j)|\mathcal{C}_1\mathcal{C}_2\mathbf{K}_2(j)(\mathbf{M_1})_{j+1}^b\mathbf{Z}^j) \\
\stackrel{(d)} = & H(\mathbf{M_1}(j)|\mathcal{C}_1\mathcal{C}_2\mathbf{K}_2(j))-H(\mathbf{M_1}(j)|\mathcal{C}_1\mathcal{C}_2\mathbf{K}_2(j)\mathbf{Z}^j) \\
= & I(\mathbf{M_1}(j);\mathbf{Z}^j|\mathcal{C}_1\mathcal{C}_2\mathbf{K}_2(j))
\end{align*}	
where $(d)$ follows that $(\mathbf{M_1})_{j+1}^b$ is independent of $\mathbf{M_1}(j)$ and $\mathbf{Z}^j$.	

Hence, 
\begin{align}
& I(\mathbf{M_1}(j);\mathbf{Z}^j|\mathcal{C}_1\mathcal{C}_2\mathbf{K}_2(j)) \nonumber \\
=& I(\mathbf{M}_{1u}(j)\mathbf{M}_{1s}(j);\mathbf{Z}^j|\mathcal{C}_1\mathcal{C}_2\mathbf{K}_2(j)) \nonumber \\
=& I(\mathbf{M}_{1u}(j)\mathbf{M}_{1s}(j);\mathbf{Z}^{j-1}|\mathcal{C}_1\mathcal{C}_2\mathbf{K}_2(j))+I(\mathbf{M}_{1u}(j)\mathbf{M}_{1s}(j);\mathbf{Z}(j)|\mathcal{C}_1\mathcal{C}_2\mathbf{K}_2(j)\mathbf{Z}^{j-1}) \nonumber \\
\stackrel{(a)}=& I(\mathbf{M}_{1u}(j)\mathbf{M}_{1s}(j);\mathbf{Z}(j)|\mathcal{C}_1\mathcal{C}_2\mathbf{K}_2(j)\mathbf{Z}^{j-1}) \nonumber  \\
=& I(\mathbf{M}_{1s}(j);\mathbf{Z}(j)|\mathcal{C}_1\mathcal{C}_2\mathbf{K}_2(j)\mathbf{Z}^{j-1})+I(\mathbf{M}_{1u}(j);\mathbf{Z}(j)|\mathcal{C}_1\mathcal{C}_2\mathbf{K}_2(j)\mathbf{Z}^{j-1}\mathbf{M}_{1s}(j)) \label{Terms}	
\end{align}
where $(a)$ follows $\mathbf{M_1}(j)$ is independent of past transmissions $\mathbf{Z}^{j-1}$, given $\mathbf{K}_2(j)$.	

We bound the two terms separately in  \eqref{Terms}. 
\begin{itemize}
	\item Consider the first term of \eqref{Terms}
	\begin{align}
	& I(\mathbf{M}_{1s}(j);\mathbf{Z}(j)|\mathcal{C}_1\mathcal{C}_2\mathbf{K}_2(j)\mathbf{Z}^{j-1})  \nonumber\\
	=& I(\mathbf{M}_{1s}(j),L_1L_2;\mathbf{Z}(j)|\mathcal{C}_1\mathcal{C}_2\mathbf{K}_2(j)\mathbf{Z}^{j-1})- I(L_1L_2;\mathbf{Z}(j)|\mathcal{C}_1\mathcal{C}_2\mathbf{K}_2(j)\mathbf{Z}^{j-1}\mathbf{M}_{1s}(j)) \nonumber\\
	=& I(\mathbf{M}_{1s}(j),L_1L_2;\mathbf{Z}(j)|\mathcal{C}_1\mathcal{C}_2\mathbf{K}_2(j)\mathbf{Z}^{j-1})-H(L_1L_2|\mathcal{C}_1\mathcal{C}_2\mathbf{K}_2(j)\mathbf{Z}^{j-1}\mathbf{M}_{1s}(j))+H(L_1L_2|\mathcal{C}_1\mathcal{C}_2\mathbf{K}_2(j)\mathbf{Z}^{j-1}\mathbf{M}_{1s}(j)\mathbf{Z}(j))  \nonumber\\
	\stackrel{(a)} \leq& I(U_1^n(L_1)U_2^n(L_2);\mathbf{Z}(j)|\mathcal{C}_1\mathcal{C}_2\mathbf{K}_2(j)\mathbf{Z}^{j-1})-H(L_1L_2|\mathcal{C}_1\mathcal{C}_2\mathbf{K}_2(j)\mathbf{Z}^{j-1}\mathbf{M}_{1s}(j))+H(L_1L_2|\mathcal{C}_1\mathcal{C}_2\mathbf{K}_2(j)\mathbf{M}_{1s}(j)\mathbf{Z}(j)) \label{Equ_a}\\
	=&  H(\mathbf{Z}(j)|\mathcal{C}_1\mathcal{C}_2\mathbf{K}_2(j)\mathbf{Z}^{j-1})-H(\mathbf{Z}(j)|U_1^n(L_1)U_2^n(L_2)\mathcal{C}_1\mathcal{C}_2\mathbf{K}_2(j)\mathbf{Z}^{j-1})  \nonumber\\
	& -H(L_1L_2|\mathcal{C}_1\mathcal{C}_2\mathbf{K}_2(j)\mathbf{Z}^{j-1}\mathbf{M}_{1s}(j))+H(L_1L_2|\mathcal{C}_1\mathcal{C}_2\mathbf{K}_2(j)\mathbf{M}_{1s}(j)\mathbf{Z}(j))  \nonumber\\	
	\stackrel{(b)} 
	\leq &  H(\mathbf{Z}(j)|\mathcal{C}_1\mathcal{C}_2\mathbf{K}_2(j))-H(\mathbf{Z}(j)|U_1^n(L_1)U_2^n(L_2)\mathcal{C}_1\mathcal{C}_2\mathbf{K}_2(j))  \nonumber\\
	& -H(L_1L_2|\mathcal{C}_1\mathcal{C}_2\mathbf{K}_2(j)\mathbf{Z}^{j-1}\mathbf{M}_{1s}(j))+H(L_1L_2|\mathcal{C}_1\mathcal{C}_2\mathbf{K}_2(j)\mathbf{M}_{1s}(j)\mathbf{Z}(j))  \nonumber\\		   
	\leq & H(\mathbf{Z}(j)|\mathcal{C}_1\mathcal{C}_2) -H(\mathbf{Z}(j)|U_1^n(L_1)U_2^n(L_2)\mathcal{C}_1\mathcal{C}_2\mathbf{K}_2(j))  \nonumber\\
	& -H(L_1L_2|\mathcal{C}_1\mathcal{C}_2\mathbf{K}_2(j)\mathbf{Z}^{j-1}\mathbf{M}_{1s}(j))+H(L_1L_2|\mathcal{C}_1\mathcal{C}_2\mathbf{K}_2(j)\mathbf{M}_{1s}(j)\mathbf{Z}(j)) \nonumber\\	
	\stackrel{(c)} = & H(\mathbf{Z}(j)|\mathcal{C}_1\mathcal{C}_2) -H(\mathbf{Z}(j)|U_1^n(L_1)U_2^n(L_2)\mathcal{C}_1\mathcal{C}_2) \nonumber\\
	& -H(L_1L_2|\mathcal{C}_1\mathcal{C}_2\mathbf{K}_2(j)\mathbf{Z}^{j-1}\mathbf{M}_{1s}(j))+H(L_1L_2|\mathcal{C}_1\mathcal{C}_2\mathbf{K}_2(j)\mathbf{M}_{1s}(j)\mathbf{Z}(j)) \nonumber\\	
	=& I(\mathbf{Z}(j);U_1^n(L_1)U_2^n(L_2)|\mathcal{C}_1\mathcal{C}_2) 	
	-H(L_1L_2|\mathcal{C}_1\mathcal{C}_2\mathbf{K}_2(j)\mathbf{Z}^{j-1}\mathbf{M}_{1s}(j))+H(L_1L_2|\mathcal{C}_1\mathcal{C}_2\mathbf{K}_2(j)\mathbf{M}_{1s}(j)\mathbf{Z}(j))  \nonumber\\
	\stackrel{(d)}\leq& nI(U_1U_2;Z)-n(\overline{R}_1+\overline{R}_2-R_{1s}-R_{2k})+n(\overline{R}_1+\overline{R}_2-R_{1s}-R_{2k}-I(U_1U_2;Z)+\delta'(\varepsilon))  \nonumber\\
	=& n\delta'(\varepsilon) \nonumber
	\end{align}		
	where $(a)$ follows that $(L_1, L_2) \rightarrow (U_1^n(L_1), U_2^n(L_2)) \rightarrow \mathbf{Z}(j)$, data processing, and conditioning reduces entropy.
	$(b)$ follows that  $H(\mathbf{Z}(j)|\mathcal{C}_1\mathcal{C}_2\mathbf{K}_2(j)\mathbf{Z}^{j-1})\leq H(\mathbf{Z}(j)|\mathcal{C}_1\mathcal{C}_2\mathbf{K}_2(j))$, and 
		\begin{align*}
		& H(\mathbf{Z}(j)|U_1^n(L_1)U_2^n(L_2)\mathcal{C}_1\mathcal{C}_2\mathbf{Z}^{j-1}) \\
		=& H(\mathbf{Z}(j)|U_1^n(L_1)U_2^n(L_2)\mathcal{C}_1\mathcal{C}_2)-I(\mathbf{Z}(j); \mathbf{Z}^{j-1}|U_1^n(L_1)U_2^n(L_2)\mathcal{C}_1\mathcal{C}_2) \\
		\stackrel{(b1)} =& H(\mathbf{Z}(j)|U_1^n(L_1)U_2^n(L_2)\mathcal{C}_1\mathcal{C}_2)
		\end{align*}    
		where $(b1)$ follows from Markov chain $\mathbf{Z}^{j-1} \rightarrow (U_1^n(L_1),\ U_2^n(L_2))\rightarrow \mathbf{Z}(j)$.
		
	 $(c)$ follows from 		
		\begin{align*}
		& H(\mathbf{Z}(j)|U_1^n(L_1)U_2^n(L_2)\mathcal{C}_1\mathcal{C}_2\mathbf{K}_2(j))\\
		=& H(\mathbf{Z}(j)|U_1^n(L_1)U_2^n(L_2)\mathcal{C}_1\mathcal{C}_2)-I(Z(j);\mathbf{K}_2(j)|U_1^n(L_1)U_2^n(L_2)\mathcal{C}_1\mathcal{C}_2)\\
		\stackrel{(c1)}=& H(\mathbf{Z}(j)|U_1^n(L_1)U_2^n(L_2)\mathcal{C}_1\mathcal{C}_2)
		\end{align*}
		where $(c1)$ follows the $Z(j)$ is independent with $\mathbf{K}_2(j)$ given $U_1^n(L_1)U_2^n(L_2)$. 	
		
		$(d)$ follows from the following three terms. Firstly,
		\begin{align*}
		& I(\mathbf{Z}(j);U_1^n(L_1)U_2^n(L_2)|\mathcal{C}_1\mathcal{C}_2) \\
		\leq & n I(\mathbf{Z_i(j)};U_{1i}1(L_1)U_{2i}(L_2)|\mathcal{C}_1\mathcal{C}_2)+n\epsilon \\
		\leq & n I(Z;U_1U_2)+n\epsilon
		\end{align*}  
		
		The second terms 
		\begin{align*}
		& H(L_1L_2|\mathcal{C}_1\mathcal{C}_2\mathbf{K}_2(j)\mathbf{Z}^{j-1}\mathbf{M}_{1s}(j)) \\
		=& H(L_1L_2,\mathbf{M}_{1u}(j)\oplus \mathbf{K}_2(j-1)|\mathcal{C}_1\mathcal{C}_2\mathbf{K}_2(j)\mathbf{Z}^{j-1}\mathbf{M}_{1s}(j)) \\	
		\geq & H(L_1L_2,\mathbf{M}_{1u}(j)\oplus \mathbf{K}_2(j-1)|\mathcal{C}_1\mathcal{C}_2\mathbf{K}_2(j)\mathbf{Z}^{j-1}\mathbf{M}_{1s}(j)\mathbf{K}_2(j-1)) \\	
		=& H(\mathbf{M}_{1u}(j)|\mathcal{C}_1\mathcal{C}_2\mathbf{K}_2(j)\mathbf{Z}^{j-1}\mathbf{M}_{1s}(j)\mathbf{K}_2(j-1)) \\ & +H(L_1L_2|\mathcal{C}_1\mathcal{C}_2\mathbf{K}_2(j)\mathbf{Z}^{j-1}\mathbf{M}_{1s}(j),\mathbf{M}_{1u}(j),\mathbf{K}_2(j-1)) \\	
		\stackrel{(f_1)} =& H(\mathbf{M}_{1u}(j)|\mathcal{C}_1\mathcal{C}_2\mathbf{K}_2(j)\mathbf{Z}^{j-1}\mathbf{M}_{1s}(j)\mathbf{K}_2(j-1)) +n(\overline{R}_1+\overline{R}_2-R_{1s}-R_{1u}-R_{2k})\\
		\stackrel{(f_2)} =& H(\mathbf{M}_{1u}(j)|\mathcal{C}_1\mathcal{C}_2\mathbf{K}_2(j)\mathbf{M}_{1s}(j)\mathbf{K}_2(j-1))+n(\overline{R}_1+\overline{R}_2-R_{1s}-R_{1u}-R_{2k})\\
		=& H(\mathbf{M}_{1u}(j)|\mathcal{C}_1\mathcal{C}_2\mathbf{K}_2(j)\mathbf{M}_{1s}(j)\mathbf{K}_2(j-1))+n(\overline{R}_1+\overline{R}_2-R_{1s}-R_{1u}-R_{2k})\\
		\stackrel{(f_3)} =& n(\overline{R}_1+\overline{R}_2-R_{1s}-R_{2k})
		\end{align*}
		where $(f_1)$ follows from the coding scheme that $H(L_1L_2|\mathcal{C}_1\mathcal{C}_2\mathbf{K}_2(j)\mathbf{Z}^{j-1}\mathbf{M}_{1s}(j),\mathbf{M}_{1u}(j),\mathbf{K}_2(j-1))= n(\overline{R}_1+\overline{R}_2-R_{1s}-R_{1u}-R_{2k})$.
		$(f_2)$ follows the Markov chain  
		\begin{align*}
		(\mathbf{Z}^{j-1},\mathbf{M}_{1s}(j),\mathbf{K}_2(j)) \rightarrow (\mathbf{K}_2(j-1),\mathbf{M}_{1s}(j),\mathbf{K}_2(j)) \rightarrow (\mathbf{M}_{1u}(j)\oplus \mathbf{K}_2(j-1),\mathbf{M}_{1s}(j),\mathbf{K}_2(j))
		\end{align*}		
		$(f_3)$ follows that $\mathbf{M}_{1u}(j)$ is independent of $(\mathcal{C}_1\mathcal{C}_2,\mathbf{K}_2(j),\mathbf{M}_{1s}(j), \mathbf{K}_2(j-1))$ and uniformly distributed over $[1:2^{nR_{1u}}]$.
		
		The third term follows Lemma \ref{Lemma_R1R2} that $H(L_1L_2|\mathcal{C}_1\mathcal{C}_2\mathbf{K}_2(j)\mathbf{M}_{1s}(j)\mathbf{Z}(j))\leq n(\overline{R}_1+\overline{R}_2-R_{1s}-R_{2k}-I(U_1U_2;Z)+\delta'(\varepsilon))$, if
		$\overline{R}_1+\overline{R}_2-R_{1s}-R_{2k}>I(U_1U_2;Z)$, $\overline{R}_1-R_{1s}>I(U_1;Z)$, $\overline{R}_2-R_{2k}>I(U_2;Z)$.
	
	\item Next we bound the second term of \eqref{Terms}.  
	\begin{align*}
	& I(\mathbf{M}_{1u}(j);\mathbf{Z}(j)|\mathcal{C}_1\mathcal{C}_2\mathbf{K}_2(j)\mathbf{Z}^{j-1}\mathbf{M}_{1s}(j)) \\
	=& I(\mathbf{M}_{1u}(j)L_1L_2;\mathbf{Z}(j)|\mathcal{C}_1\mathcal{C}_2\mathbf{K}_2(j)\mathbf{Z}^{j-1}\mathbf{M}_{1s}(j))-I(L_1L_2;\mathbf{Z}(j)|\mathcal{C}_1\mathcal{C}_2\mathbf{K}_2(j)\mathbf{Z}^{j-1}\mathbf{M}_{1s}(j)\mathbf{M}_{1u}(j)) \\
	\leq& I(U_1^n(L_1)U_2^n(L_2);\mathbf{Z}(j)|\mathcal{C}_1\mathcal{C}_2\mathbf{K}_2(j)\mathbf{Z}^{j-1}\mathbf{M}_{1s}(j))-H(L_1L_2|\mathcal{C}_1\mathcal{C}_2\mathbf{K}_2(j)\mathbf{Z}^{j-1}\mathbf{M}_{1s}(j)\mathbf{M}_{1u}(j))\\
	& +H(L_1L_2|\mathcal{C}_1\mathcal{C}_2\mathbf{K}_2(j)\mathbf{Z}(j)\mathbf{M}_{1s}(j)\mathbf{M}_{1u}(j)) \\
	\leq& I(U_1^n(L_1)U_2^n(L_2);\mathbf{Z}(j)|\mathcal{C}_1\mathcal{C}_2\mathbf{K}_2(j)\mathbf{M}_{1s}(j))-H(L_1L_2|\mathcal{C}_1\mathcal{C}_2\mathbf{K}_2(j)\mathbf{Z}^{j-1}\mathbf{M}_{1s}(j)\mathbf{M}_{1u}(j)) \\
	&+H(L_1L_2|\mathcal{C}_1\mathcal{C}_2\mathbf{K}_2(j)\mathbf{Z}(j)\mathbf{M}_{1s}(j)\mathbf{M}_{1u}(j)) \\
	\stackrel{(a)} \leq& nI(U_1U_2;Z)-H(L_1L_2|\mathcal{C}_1\mathcal{C}_2\mathbf{K}_2(j)\mathbf{Z}^{j-1}\mathbf{M}_{1s}(j)\mathbf{M}_{1u}(j))+H(L_1L_2|\mathcal{C}_1\mathcal{C}_2\mathbf{K}_2(j)\mathbf{Z}(j)\mathbf{M}_{1s}(j)\mathbf{M}_{1u}(j)) \\
	\stackrel{(b)} \leq& nI(U_1U_2;Z)-H(L_1L_2|\mathcal{C}_1\mathcal{C}_2\mathbf{K}_2(j)\mathbf{Z}^{j-1}\mathbf{M}_{1s}(j)\mathbf{M}_{1u}(j))+n(\overline{R}_1+\overline{R}_2-R_{1s}-R_{1u}-R_{2k}-I(U_1U_2;Z)+\delta'(\varepsilon))\\
	=& n(\overline{R}_1+\overline{R}_2-R_{1s}-R_{1u}-R_{2k}))-H(L_1L_2|\mathcal{C}_1\mathcal{C}_2\mathbf{K}_2(j)\mathbf{Z}^{j-1}\mathbf{M}_{1s}(j)\mathbf{M}_{1u}(j))+n\delta'(\varepsilon) \\
	\stackrel{(c)} \leq& n(\overline{R}_1+\overline{R}_2-R_{1s}-R_{1u}-R_{2k}))-n(\overline{R}_1+\overline{R}_2-R_{1s}-R_{1u}-R_{2k}-\delta'''(\varepsilon)) \\
	=& n(\delta'''(\varepsilon))
	\end{align*}
	where $(a)$ follows the same analysis if \eqref{Equ_a}; $(b)$ follows Lemma \ref{Lemma_R1R2}  that $H(L_1L_2|\mathcal{C}_1\mathcal{C}_2\mathbf{K}_2(j)\mathbf{Z}(j)\mathbf{M}_{1s}(j)\mathbf{M}_{1u}(j)) \leq n(\overline{R}_1+\overline{R}_2-R_{1s}-R_{1u}-R_{2k}-I(U_1U_2;Z)+\delta'(\varepsilon))$, if $\overline{R}_1+\overline{R}_2-R_{1s}-R_{2k}-R_{1u}\geq I(U_1U_2;Z)$, $\overline{R}_1-R_{1s}-R_{1u}\geq I(U_1;Z)$, $\overline{R}_2-R_{2k}\geq I(U_2;Z)$.
	
	$(c)$ follows that
		\begin{align*}
		H(L_1L_2|\mathcal{C}_1\mathcal{C}_2\mathbf{K}_2(j)\mathbf{Z}^{j-1}\mathbf{M}_{1s}(j)\mathbf{M}_{1u}(j)) = n(\overline{R}_1+\overline{R}_2-R_{1s}-R_{2k}-R_{1u}) 
		\end{align*}
\end{itemize}

\subsection{Rate Analysis}
Consider all the conditions are fulfilled to keep confidential messages transmission.  
\begin{align*}
& \overline{R}_1=R_{1u}+R_{1s}+R_{1x}  \\
& \overline{R}_2=R_{2}+R_{2k}+R_{2x} \\
& \overline{R}_{1s}=R_{1u}+R_{1s} \\
& R_{1u}\leq R_{2k} \\
& R_{2}+R_{2k}+R_{2x} \leq I(U_2;Y_1|X_1,Q) \\
& R_{1u}+R_{1s}+R_{1x} \leq I(U_1;Y_2|X_2,Q) \\
& R_{1x}+R_{2}+R_{2x}\geq I(U_1U_2;Z) \\
& R_{1x} \geq  I(U_1;Z)    \\	
& R_{2}+R_{2x}\geq I(U_2;Z) 
\end{align*}
After the Fourier-Motzkin elimination, the achievable secrecy rate region are the union of non-negative rate pairs $(\overline{R}_{1s}, R_2)$ satisfying
\begin{align*}
& \overline{R}_{1s}\leq I(U_1;Y_2|X_2,Q)+I(U_2;Y_1|X_1,Q)-I(U_1U_2;Z) \\
& \overline{R}_{1s}\leq I(U_1;Y_2|X_2,Q)-I(U_1;Z)\\ 
& R_2\leq I(U_2;Y_1|X_1,Q)     
\end{align*} 
over all $p(q)p(u_1|q)p(u_2|q)p(x_1|u_1)p(x_2|u_2)$, when $I(U_2;Y_1|X_1,Q)\geq I(U_2;Z)$.

\section{Proof of Theorem \ref{Thm_OuterBound} } \label{Proof_Outer}
\begin{IEEEproof}
	
	First, we consider $R_2.$
	\begin{align*}
	nR_2=& H(M_2)=H(M_2|M_{1}) \\
	=& I(M_2; Y_1^n|M_{1},X_1^n)+H(M_2|M_{1}, Y_1^n,X_1^n)\\
	\stackrel{(a)}{\leq} & I(M_2; Y_1^n|M_{1},X_1^n)+n\epsilon\\
	=& \sum_{i=1}^{n} I(M_2; Y_{1i}|M_{1}, Y_{1}^{i-1},X_1^n)+n\epsilon\\
	\leq & \sum_{i=1}^{n} I(X_1^{i-1}, X_{1,i+1}^n, X_{2i}, M_2, Y_{1}^{i-1}; Y_{1i}|  X_{1i})+n\epsilon\\
	= & \sum_{i=1}^{n} I(X_{2i}; Y_{1i}|  X_{1i})+I(X_1^{i-1}, X_{1,i+1}^n, M_2, Y_{1}^{i-1}; Y_{1i}|X_{2i}, X_{1i})+n\epsilon\\
	= &\sum_{i=1}^{n} I(X_{2i}; Y_{1i}|  X_{1i})+n\epsilon\\
	= & nI(X_{2}; Y_{1}|  X_{1})+n\epsilon.
	\end{align*}
	Now We define the following auxiliary random variables to proceed to $\overline{R}_{1s}.$
	\begin{align}\label{Def_UV}
	U_i=X_2^{i-1} Y_2^{i-1} Z_{i+1}^n, V_i=(W_1, U_i)
	\end{align}
	\begin{align}
	\overline{R}_{1s}=& H(M_{1}) \leq H(M_{1}|Z^n)+n\epsilon \nonumber\\
	=& H(M_{1}|Z^n)-H(M_{1}|Y_2^n, X_2^n)+H(M_{1}|Y_2^n, X_2^n)\nonumber\\
	= & H(M_{1})-I(M_{1};Z^n)-H(M_{1})+I(M_{1};Y_2^n, X_2^n)\nonumber\\
	&+H(M_{1}|Y_2^n X_2^n)\nonumber\\
	=&-I(M_{1};Z^n)+I(M_{1};Y_2^n, X_2^n)+H(M_{1}|Y_2^n, X_2^n)\nonumber\\
	\stackrel{(a)}\leq& I(M_{1};X_2^n, Y_2^n)-I(M_{1};Z^n)+n\delta_n\nonumber\\
	=& \sum_{i=1}^n [I(M_{1}; X_{2i}, Y_{2i}|X_2^{i-1}, Y_2^{i-1})
	-I(M_{1};Z_i|Z_{i+1}^n)]+n\delta_n\nonumber\\
	=& \sum_{i=1}^n [I(M_{1},Z_{i+1}^n; X_{2i}, Y_{2i}|X_2^{i-1}, Y_2^{i-1})
	-I(Z_{i+1}^n; X_{2i}, Y_{2i}|X_2^{i-1}, Y_2^{i-1}, M_{1})\nonumber\\
	&-I(M_{1},X_2^{i-1}, Y_2^{i-1};Z_i|Z_{i+1}^n)
	+I(X_2^{i-1}, Y_2^{i-1};Z_i|Z_{i+1}^n, M_{1})]+n\delta_n\nonumber\\
	\stackrel{(b)}=& \sum_{i=1}^n [I(M_{1}, Z_{i+1}^n; X_{2i},  Y_{2i}|X_2^{i-1}, Y_2^{i-1})-I(M_{1}, X_2^{i-1}, Y_2^{i-1};Z_i|Z_{i+1}^n)]+n\delta_n\nonumber\\
	=& \sum_{i=1}^n [I(Z_{i+1}^n; X_{2i}, Y_{2i}|X_2^{i-1},  Y_2^{i-1})+I(M_{1}; X_{2i}, Y_{2i}|X_2^{i-1}, Y_2^{i-1}, Z_{i+1}^n)\nonumber\\
	&-I(X_2^{i-1}, Y_2^{i-1};Z_i|Z_{i+1}^n)\nonumber-I(M_{1};Z_i|X_2^{i-1}, Y_2^{i-1} Z_{i+1}^n)]+n\delta_n\nonumber\\
	\stackrel{(b)}=& \sum_{i=1}^n [I(M_{1}; X_{2i}, Y_{2i}|X_2^{i-1}, Y_2^{i-1}, Z_{i+1}^n)-I(M_{1};Z_i|X_2^{i-1}, Y_2^{i-1},  Z_{i+1}^n)]+n\delta_n \label{R_1e1}\\
	\stackrel{(c)}=& \sum_{i=1}^n [I(V_i; X_{2i},  Y_{2i}|U_i)-I(V_i;Z_i|U_i)]+n\delta_n\nonumber\\
	\stackrel{(d)}= & \sum_{i=1}^n [I(V; X_{2},  Y_{2}|U)-I(V;Z|U)]+n\delta_n\nonumber\\
	=& n[I(V; X_{2},  Y_{2}|U)-I(V;Z|U)]+n\delta_n   \label{Equ_Conv_R1ePf}
	\end{align}
	where
	$(a)$ follows by the Fano's inequality;
	$(b)$ follows from the Csisz\'{a}r sum identity \cite{el2013achievable};
	$(c)$ follows by defining $U_i=(X_2^{i-1}, Y_2^{i-1}, Z_{i+1}^n)$ and $V_i=(M_{1}, U_i)$ in \eqref{Def_UV}. 
	and $(d)$ is by introducing a time-sharing random variable $Q,$ and define $U=(U_Q, Q)$ and $V=(V_Q,Q), X_{2}=X_{2Q}, Y_{2}=Y_{2Q}, Z=Z_Q$,  then
	\begin{align*}
	&I(V; X_{2} Y_{2}|U)-I(V;Z|U)\\
	=&I(V_G,G; X_{2G} Y_{2G}|U_G,G)-I(V_G,G;Z_G|U_G,G)\\
	=&I(V_G; X_{2G} Y_{2G}|U_G,G)+I(G; X_{2G} Y_{2G}|U_G,V_G,G)\\
	&-I(V_G;Z_G|U_G,G)-I(G;Z_G|U_G,V_G,G)\\
	= & I(V_G; X_{2G} Y_{2G}|U_G,G)-I(V_G;Z_G|U_G,G)
	\end{align*}
	The equivocation rate \eqref{Equ_Conv_R1ePf}
	\begin{align}
	\overline{R}_{1s}\leq & \sum_{i=1}^n [I(V_i; X_{2i} Y_{2i}|U_i)-I(V_i;Z_i|U_i)]+n\delta_n\nonumber\\
	=& n[I(V_G; X_{2G} Y_{2G}|U_G,G)-I(V_G;Z_G|U_G,G)]+n\delta_n \nonumber\\
	\leq & n[I(V; X_{2}Y_{2}| U)-I(V;Z|U)]+n \delta_n \label{Equ_R1e_Bound}
	\end{align}
\end{IEEEproof} 
\renewcommand\refname{reference}
\bibliographystyle{IEEEtran}
\bibliography{TwoWay}
\end{document}